\documentclass[final,5p,times,twocolumn]{elsarticle}

\usepackage{amssymb}
\usepackage{amsmath}
\usepackage{siunitx}
\usepackage{comment}
\usepackage{xcolor}
\usepackage{hyperref}
\usepackage{cleveref}

\biboptions{sort&compress}

\journal{Acta Materialia}

\begin{document}


\begin{frontmatter}

\title{Response-function-optimized phase field modeling of solute trapping and solute drag in rapid alloy solidification}

\author[label1,label2]{Joni Kaipainen\corref{cor1}}
\cortext[cor1]{Corresponding author.}
\ead{joni.kaipainen@vtt.fi}

\author[label1]{Tatu Pinomaa}
\author[label2]{Nikolas Provatas}

\affiliation[label1]{organization={Integrated Computational Materials Engineering group, VTT Technical Research
Centre of Finland Ltd.},
             addressline={Maarintie 3},
             city={Espoo},
             postcode={02150},
             country={Finland}}

\affiliation[label2]{organization={Centre for the Physics of Materials, Department of Physics, McGill University},
             addressline={3600 Rue University},
             city={Montreal},
             postcode={H1M 2N4},
             state={QC},
             country={Canada}}

\begin{abstract}
Quantitative prediction of rapid-solidification microstructures requires phase field models that represent the velocity dependence of interfacial properties, including solute partitioning, kinetic liquidus response, solute drag, and kinetic undercooling. These response functions control both microsegregation and morphology selection, but are difficult to prescribe accurately in phase field simulations that employ enlarged interfaces for numerical efficiency. We introduce an optimization-based calibration strategy that embeds target sharp-interface response functions into a dilute alloy phase field formulation by treating the interfacial diffusivity interpolation function as a response-matching degree of freedom. The optimized diffusivity functions are obtained from one-dimensional steady-state phase field solutions constrained to reproduce prescribed continuous-growth-model targets for velocity-dependent solute trapping and drag-modified liquidus kinetics. We demonstrate the accuracy and versatility of the calibrated model in dilute Al--Cu by reproducing the prescribed response functions for intermediate solute drag coefficients relevant to rapid solidification. Two-dimensional simulations are  {carried out over a range of drag coefficients and pulling velocities relevant to additive manufacturing, demonstrating} that quantitative changes in the prescribed interface kinetics can systematically shift the morphology from cellular to banded growth. Additionally, we extend the formulation to dilute multicomponent alloys, enabling independent specification of equilibrium partition coefficients and liquidus slopes for multiple solute species. The framework separates the physical specification of nonequilibrium interface kinetics from their diffuse-interface realization and provides a practical route for embedding response functions into quantitative phase field computations.
\end{abstract}


\end{frontmatter}

\section{Introduction}
\label{sec:introduction}

Modern manufacturing processes such as laser powder bed fusion, directed energy deposition, welding, and laser surface remelting involve localized melting and rapid resolidification under steep thermal gradients, high cooling rates, and interface velocities that can vary strongly within a single melt pool~\cite{Herzog2016_AM_review,Tang2022_meltpool,nanotoday,Chang2026_meltpool}. These conditions can produce refined microstructures, extended solid solubility, metastable solidification pathways, and nonequilibrium structures such as bands. However, they also make solute segregation, cellular or dendritic spacings, and texture selection sensitive to alloy chemistry and thermal history~\cite{McKeown2016_banding,Gianoglio2020_banding,Ji2024_banding,Zhong2025_nature,nanotoday}. Predictive simulation tools are therefore essential to connect controllable processing conditions to the microstructures generated during rapid solidification.

A central difficulty in modeling rapid solidification is that the solid-liquid interface is driven far from local equilibrium. In this regime, equilibrium partition coefficients and equilibrium liquidus slopes no longer provide an accurate description of the local conditions at the moving interface. Instead, solute partitioning and kinetic undercooling become velocity dependent, and the resulting nonequilibrium interface response can influence whether the interface evolves toward planar, cellular, dendritic, or oscillatory growth~\cite{KGT1986,KarmaSarkissian1993,Tourret2023}. Binary alloys provide the clearest setting for defining and validating these response functions. However, many relevant engineering alloys such as steels~\cite{Pinomaa2020_316L} and high entropy alloys~\cite{George2019_hea_review,npj_surrogate} contain multiple solute species, each of which may redistribute and trap over a different velocity range. A general rapid solidification phase field formulation should therefore be able to accommodate multiple components while reducing to the binary case in the appropriate limit.

A widely used sharp-interface description of nonequilibrium solute redistribution is provided by the continuous growth model (CGM) and its extensions to solute drag and multicomponent alloys~\cite{Aziz1982,Aziz1988,AzizBoettinger1994_drag,Ludwig1998,Tourret2023}. In this framework, solute trapping results from the competition between interface migration and short-range solute redistribution across the solid-liquid interface. Throughout this work, \(k_i\) denotes the equilibrium partition coefficient of solute \(i\), while \(k_i(V)\) denotes its corresponding velocity-dependent nonequilibrium partition coefficient. For a dilute alloy containing \(n_s\) solute species, each solute is assigned its own CGM response form,
\begin{equation}
    k_i(V)
    =
    \frac{k_i + V/V_{D,i}}
         {1 + V/V_{D,i}},
    \qquad
    i=1,\ldots,n_s,
    \label{eq:intro_cgm_ki}
\end{equation}
where \(V_{D,i}\) is the corresponding diffusive speed for redistribution of solute $i$ across the moving interface~\cite{Ludwig1998,Tourret2023}. It is easy to see that \(k_i(V)\to k_i\) for \(V\ll V_{D,i}\), whereas \(k_i(V)\to 1\) as the interface velocity becomes large compared with the redistribution speed of that species. The quantities \(V_{D,i}\) are interface-kinetic material parameters rather than numerical phase field parameters that may be inferred from, e.g., rapid solidification experiments~\cite{SmithAziz1994_Al,Kittl2000_SiAs_drag} or atomistic simulations of moving solid-liquid interfaces~\cite{Yang2011_drag_MD,Haapalehto2022}. For \(n_s=1\), Eq.~\eqref{eq:intro_cgm_ki} reduces to the usual binary CGM partition coefficient~\cite{Aziz1982,Aziz1988}.

The partition coefficients alone do not fully specify the nonequilibrium interface response, because the interface velocity also depends on the thermodynamic driving force available for crystallization. The idea of solute redistribution dragging an interface and reducing the effective driving force can be traced back to the work of Cahn on grain boundaries~\cite{Cahn1962}. In alloy solidification, the free energy decrease associated with transforming liquid into solid may be divided between crystallization and solute redistribution~\cite{AzizBoettinger1994_drag,Pinomaa_Laukkanen_Provatas_2020}. If \(\Delta G_{\rm DF}\) denotes the total driving force and \(\Delta G_{\rm D}\) the contribution dissipated by solute redistribution, the effective driving force available for interface motion may be represented schematically as
\begin{equation}
    \Delta G_{\rm eff}
    =
    \Delta G_{\rm DF}
    -
    \alpha \Delta G_{\rm D},
    \label{eq:intro_effective_driving_force}
\end{equation}
where \(\alpha\) is a solute drag coefficient~\cite{AzizBoettinger1994_drag}. The limit \(\alpha=0\) corresponds to no solute drag, for which the full driving free energy is available for interface motion, whereas \(\alpha=1\) corresponds to full subtraction of the redistribution contribution from the driving force. Intermediate values represent partial solute drag, which is supported by theoretical treatments and atomistic simulations of alloy solidification~\cite{Ahmad1998,Yang2011_drag_MD,Kavousi2020_drag,Haapalehto2022,Hareland2022,Antillon2023}. In this interpretation, solute drag is not an independent correction to the phase diagram; instead, it represents the reduction of the effective driving force caused by dissipative solute redistribution at the moving interface.

In the thermodynamic formulation underlying Eq.~\eqref{eq:intro_effective_driving_force}, the crystallization and redistribution contributions are evaluated from the solid and liquid compositions and chemical potentials at the interface, and the effective driving force determines the interface velocity through a collision-limited growth relation~\cite{AzizBoettinger1994_drag}. For dilute alloys, evaluating this relation gives the velocity-dependent kinetic liquidus slopes. Using the positive magnitude \(m_i\) of the equilibrium liquidus slope associated with solute \(i\), the multicomponent response of the liquidus of solute \(i\) is written as
\begin{equation}
    \frac{m_i(V;\alpha_i)}{m_i}
    =
    \frac{
    1-k_i(V)
    +
    \big[
        k_i(V)
        +
        \alpha_i\left(1-k_i(V)\right)
    \big]
    \ln\!\left[k_i(V)/k_i\right]
    }{
    1-k_i
    },
    \label{eq:intro_cgm_mi}
\end{equation}
for \(i=1,\ldots,n_s\). Equation~\eqref{eq:intro_cgm_mi} reduces to the Aziz--Boettinger binary expression for a single solute~\cite{AzizBoettinger1994_drag}. Within the dilute independent-solute approximation, each solute may be assigned its own \(k_i\), \(V_{D,i}\), \(m_i\), and phenomenological drag coefficient \(\alpha_i\). In the present calculations, species-resolved drag information is unavailable, and we therefore adopt the common-drag case \(\alpha_i=\alpha\) for all solutes.

Equations~\eqref{eq:intro_cgm_ki} and \eqref{eq:intro_cgm_mi} define the sharp-interface response functions used below as optimization targets. The role of a phase field model here is not to determine \(V_{D,i}\), \(k_i(V)\), \(m_i(V;\alpha)\), or \(\alpha\) from first principles, but  to accurately emulate a controlled, diffuse-interface, realization of these prescribed nonequilibrium response functions. These target functions may be obtained from theoretical work~\cite{Ahmad1998}, rapid solidification experiments~\cite{SmithAziz1994_Al,Kittl2000_SiAs_drag}, or atomistic simulations~\cite{Yang2011_drag_MD,Kavousi2020_drag,Haapalehto2022}. This separation between physical response-function specification and diffuse-interface numerical realization is central to the optimization strategy developed in this work. Reproducing both \(k_i(V)\) and \(m_i(V;\alpha)\) is important because solute trapping and solute drag affect different aspects of the moving-interface response: the former controls nonequilibrium solute partitioning, while the latter modifies the velocity-dependent thermodynamic driving force available for crystallization. The consequences of this combined response for microstructure selection are demonstrated in this work.

Mesoscale solidification simulations provide valuable tools because they connect interface-scale kinetics to microstructural features that can be measured and controlled, such as segregation patterns, grain competition, cellular or dendritic spacings, interface stability, and band formation~\cite{Tourret2015,Wang2022_segregation_spacing,Ji2023,Ji2025,Zhong2025_nature}. Phase field models are particularly well suited to this task because they describe the solid-liquid boundary as a smoothly varying field rather than an explicitly tracked surface, allowing planar, cellular, dendritic, and oscillatory growth modes to emerge within a single computational framework. Quantitative phase field models have been highly successful for near-equilibrium alloy solidification, where thin-interface analysis and anti-trapping corrections remove nonequilibrium effects to recover local-equilibrium boundary conditions while using computational interface widths much larger than that of the physical solid-liquid interface~\cite{Echebarria2004,Tourret2015,book}. 

Rapid solidification poses a different challenge. In this regime, solute trapping and kinetic undercooling are not numerical artifacts to be eliminated, but physical nonequilibrium effects that must be retained quantitatively~\cite{Pinomaa_Laukkanen_Provatas_2020,Tourret2023}. At the same time, tractable phase field simulations on experimentally relevant microstructural length scales still require a diffuse-interface thickness \(W\) that is larger than the physical interface width \(W_0\). Without suitable corrections defined for this regime, upscaling $W$ changes the effective solute redistribution kinetics and kinetic undercooling away from that defined by Eqs.~\eqref{eq:intro_cgm_ki} and \eqref{eq:intro_cgm_mi}. Earlier one-dimensional phase field studies had already shown that nonequilibrium solute redistribution can be  represented in the sharp-interface limit of phase field models; Wheeler \textit{et al.} introduced an isothermal binary alloy model that reduced segregation with increasing interface velocity and approached partitionless solidification at high velocity~\cite{Wheeler1993_PF_trapping}, and related work later examined solute trapping and solute drag during rapid solidification~\cite{Ahmad1998,Danilov2007_pf_trapping}. The challenge for mesoscale rapid solidification simulations, however, is not simply to produce trapping, but to emulate quantitatively prescribed response functions while using computationally enlarged interfaces.

Recent far-from-equilibrium dilute binary alloy phase field formulations address this issue using two complementary approaches. Pinomaa and Provatas~\cite{Pinomaa2019} extended the commonly used anti-trapping model of Echebarria \textit{et al.}~\cite{Echebarria2004} to quantitatively predict the chemical potential jump across the interface for moderately rapid solidification rates. This made, for the first time, quantitative phase field predictions of rapid solidification morphologies feasible in regimes where local-equilibrium assumptions begin to break down. Building on the same modified anti-trapping strategy, Kavousi and Asle Zaeem subsequently introduced an alternative parameterization designed to reproduce CGM trapping kinetics with reduced sensitivity to the diffuse-interface width~\cite{Kavousi2021}. More recently, Ji \textit{et al.}~\cite{Ji2023,Ji2025} developed a methodology for controlling spurious trapping at very rapid solidification rates for computational solid-liquid interface thicknesses \(W\) greater than the physical one \(W_0\) by enhancing the solute diffusivity inside the diffuse interface as
\begin{equation}
    D_i(\phi)=D_l^i q_i(\phi),
    \label{eq:intro_diffusivity_interpolation}
\end{equation}
where \(D_l^i\) is the liquid diffusivity of solute \(i\), and \(q_i(\phi)\) is an interfacial diffusivity interpolation function. In the binary alloy case, a single function \(q(\phi)\) is tuned to compensate for the reduced effective redistribution speed caused by the enlarged interface width, while preserving the response functions \(k(V)\) and \(m(V)\) prescribed by the model interpolation functions at \(W=W_0\)~\cite{Ji2025}. This strategy has made it possible to reproduce CGM-like solute trapping behavior over a broad range of interface velocities and to simulate far-from-equilibrium microstructure formation, including banded structures and absolute stability limits, on microstructural length scales~\cite{Tourret2024,Ji2025,Mancias2025}. A similar idea has also been utilized by Li \textit{et al.}~\cite{Li2024_kks,Li2025_interpolation}, building on the Kim-Kim-Suzuki model~\cite{Kim1999} and the finite-interface dissipation model introduced in Ref.~\cite{Steinbach2012}. These implementations employ simple prescribed forms of \(q(\phi)\), typically with a single adjustable parameter. Such forms can be remarkably effective, but they also restrict the set of attainable response functions and do not provide a systematic way to balance accuracy against smoothness of the interfacial diffusivity profile.

In this work, we formulate the selection of the interfacial diffusivity functions \(q_i(\phi)\) as an inverse response-matching problem. Each \(q_i(\phi)\) is represented by a smooth endpoint-preserving expansion and optimized so that one-dimensional steady-state planar-interface solutions reproduce prescribed target functions \(k_{i,\rm tar}(V)\) and \(m_{i,\rm tar}(V)\), as given in Eqs.~\eqref{eq:intro_cgm_ki} and~\eqref{eq:intro_cgm_mi}. This makes the calibration of the enlarged-interface phase field model explicit and tolerance-controlled: the interpolation is selected according to the quantitative accuracy with which the resulting diffuse-interface solutions reproduce the desired nonequilibrium partitioning and kinetic liquidus responses over the velocity range of interest. The endpoint constraints ensure that the bulk solid and liquid diffusivities are unchanged, while the optimization modifies properties only within the diffuse-interface region. A regularization term is then used to select the smoothest interpolation that satisfies prescribed accuracy tolerances for both response functions.

To extend this response-matching strategy beyond binary alloys, we also introduce a dilute multicomponent phase field model motivated by the component-specific susceptibilities used in traditional parabolic free-energy constructions~\cite{Aagesen2018_grand_potential,book}. Retaining this freedom within the present dilute thermodynamic formulation permits the independent specification of equilibrium pairs \((k_i,m_i)\) for each solute, thereby avoiding the nonphysical relation between partition coefficients and liquidus slopes that arises from a direct multicomponent extension of the ideal-dilute free energy with a universal thermodynamic prefactor. This formulation reduces to the binary model of Refs.~\cite{Ji2023,Ji2025} when only one solute is present. The framework developed separates the physical specification of nonequilibrium interface kinetics from their numerical realization in the large $W$ limit of the phase field model: the target functions may come from CGM theory, experiment, atomistic simulation, or inverse calibration, while the optimized \(q_i(\phi)\) allow for controlled diffuse-interface representation of the target kinetics.

In what follows, we first formulate the phase field model, including its extension from binary to dilute multicomponent alloys, and then introduce the response-function matching framework used to calibrate the interfacial diffusivity functions. We validate the optimization strategy in binary Al--Cu by matching CGM target functions with different solute drag coefficients. The significance of quantitatively embedding these responses is then examined  in two-dimensional directional-solidification simulations spanning several pulling velocities and drag coefficients. The resulting transition between cellular and oscillatory banded growth {demonstrates} how differences in the prescribed kinetic liquidus response can propagate into multidimensional phase field predictions. We complement these simulations with a reduced analysis of the corresponding mean-front response. We finally present a case study with the multicomponent formulation in a dilute Al--Si--Cu ternary alloy to demonstrate how the same response-function matching strategy extends beyond binary alloys.

\section{Dilute multicomponent phase field formulation}
\label{sec:model_development}

Motivated by the binary far-from-equilibrium formulation of Ji \textit{et al.}~\cite{Ji2023,Ji2025}, we formulate a multicomponent phase field model that can assign to each dilute solute species an independent equilibrium partition coefficient and liquidus slope. This section establishes the thermodynamic construction, evolution equations, and planar-interface response functions that define the model, and its calibration to emulate the CGM model is presented in Sec.~\ref{sec:q_optimization}. 

A direct extension to multiple components of the ideal dilute Helmholtz free-energy density used in  phase field models becomes restrictive if all solutes share the same common thermodynamic prefactor \(A_0=RT_m/v_0\), where \(R\) is the gas constant, \(T_m\) is the melting point of the pure solvent and \(v_0\) is the molar volume. In that case, the equilibrium partition coefficients fix the solute free-energy jumps, and the resulting liquidus hypersurface imposes a constant ratio \(m_i/(1-k_i)\) for all solute species. The derivation of this van't Hoff restriction from the equilibrium liquidus hypersurface, together with its appearance in the Gibbs--Thomson solvability condition, is given in Supplementary Section~S1.

This restriction is not inherent to a Helmholtz description itself. It arises from using a universal concentration susceptibility for all solutes. We therefore introduce component-specific local susceptibilities and then choose a functional form for the diagonal dilute limit. The model derivation is conveniently written using a grand-potential representation because, in this form, concentrations and susceptibilities follow directly from derivatives of the grand-potential density of a phase with respect to diffusion potentials~\cite{Plapp2011,book}. The final equations are nevertheless written in concentration variables, since the diagonal dilute relation between \(c_i\) and \(\mu_i\) is algebraically invertible. The detailed derivation, including the ideal dilute Helmholtz restriction and the equivalent susceptibility-based construction, is given in Supplementary Material and summarized below.

We consider one solvent and \(n_s\) solute species indexed \(i=1,\ldots,n_s\). In a general multicomponent grand-potential description, each bulk phase \(\alpha=s,l\) is described by a phase grand-potential density
\begin{equation}
    \omega^\alpha
    =
    \omega^\alpha(\boldsymbol{\mu},T),
\end{equation}
where
\begin{equation}
    \boldsymbol{\mu}
    =
    (\mu_1,\ldots,\mu_{n_s})
\end{equation}
are interdiffusion potentials. The corresponding phase concentrations are
\begin{equation}
    c_i^\alpha(\boldsymbol{\mu},T)
    =
    -\frac{\partial \omega^\alpha}{\partial \mu_i},
    \label{eq:general_concentration_from_gp}
\end{equation}
and the susceptibility matrix is
\begin{equation}
    X_{ij}^\alpha
    =
    \frac{\partial c_i^\alpha}{\partial \mu_j}
    =
    -\frac{\partial^2 \omega^\alpha}
    {\partial \mu_i \partial \mu_j}.
    \label{eq:general_susceptibility_matrix}
\end{equation}
In a fully coupled multicomponent thermodynamic model, \(X_{ij}^\alpha\) is generally dense, and the diffusion equations are most naturally written in terms of diffusion potentials and a mobility matrix~\cite{book}. In the present work, we restrict attention to the dilute independent-solute approximation, in which solute-solute thermodynamic couplings and cross mobilities are neglected. Thus,
\begin{equation}
    X_{ij}^\alpha
    \approx
    X_i^\alpha \delta_{ij}. 
    \label{eq:diagonal_susceptibility_approx}
\end{equation}
This diagonalization further allows for the simple construction of evolution equations in terms of concentration.

\subsection{Diagonal dilute thermodynamics}
\label{subsec:diagonal_dilute_thermo}

In the diagonal non-interacting dilute limit, the phase grand-potential density is taken to be separable in the solute diffusion potentials,
\begin{equation}
    \omega^\alpha(\boldsymbol{\mu},T)
    =
    \omega_0^\alpha(T)
    -
    \sum_{i=1}^{n_s}
    \mathcal A_i
    \exp\left[
        \frac{\mu_i-\epsilon_i^\alpha}{\mathcal A_i}
    \right],
    \label{eq:phase_gp_diagonal_dilute}
\end{equation}
where \(\omega_0^\alpha(T)\) is the pure-solvent contribution and \(\epsilon_i^\alpha\) is the dilute solute free-energy offset in phase \(\alpha\), and $\mathcal{A}_i$ is a susceptibility parameter for phase $\alpha$ with respect to solute $i$. Equation~\eqref{eq:phase_gp_diagonal_dilute} gives
\begin{equation}
    c_i^\alpha
    =
    \exp\left[
        \frac{\mu_i-\epsilon_i^\alpha}{\mathcal A_i}
    \right],
    \label{eq:phase_concentration_diagonal_dilute}
\end{equation}
and therefore
\begin{equation}
    X_i^\alpha
    =
    \frac{\partial c_i^\alpha}{\partial\mu_i}
    =
    \frac{c_i^\alpha}{\mathcal A_i}.
    \label{eq:phase_susceptibility_diagonal_dilute}
\end{equation}

The equivalence between the local grand-potential and Helmholtz representations can be made explicit through the corresponding Legendre transformation~\cite{Zhang2025}. Using \(f^\alpha=\omega^\alpha+\sum_i\mu_i c_i^\alpha\) gives
\begin{equation}
    f^\alpha
    =
    \omega^\alpha_0(T)
    +
    \sum_i
    \left[
        \mathcal A_i
        \left(c_i^\alpha\ln c_i^\alpha-c_i^\alpha\right)
        +\epsilon_i^\alpha c_i^\alpha
    \right],
    \label{eq:dilute_helmholtz_density}
\end{equation}
and therefore
\begin{equation}
    \frac{\partial^2 f^\alpha}{\partial(c_i^\alpha)^2}
    =
    \frac{\mathcal A_i}{c_i^\alpha}
    =
    \left(X_i^\alpha\right)^{-1}.
    \label{eq:dilute_inverse_susceptibility}
\end{equation}
Therefore \(\mathcal A_i/c_i^\alpha\) plays the role of the local curvature, or inverse susceptibility, that appears in commonly used parabolic phase free-energy approximations~\cite{book,Aagesen2018_grand_potential}. The present formulation retains the logarithmic dilute form, as in grand-potential formulations that also treat dilute-solution free energies explicitly~\cite{Plapp2011,Aagesen2018_grand_potential}, rather than replacing it by its quadratic expansion. In the strictly ideal dilute Helmholtz form commonly used in the literature, however, the logarithmic prefactor is fixed by temperature and molar volume, corresponding here to a common value \(\mathcal A_i=A_0\) for all solute species. In contrast, we retain \(\mathcal A_i\) as a component-specific susceptibility parameter and fix it below by matching to the prescribed dilute liquidus slope.

To proceed, we define a phase field variable that satisfies \(\phi=+1\) in the bulk solid phase and \(\phi=-1\) in the bulk liquid phase.
Across the diffuse interface, the solute free-energy offset is interpolated as
\begin{equation}
    \epsilon_i(\phi)
    =
    \bar{\epsilon}_i
    +
    \frac{\Delta\epsilon_i}{2}g(\phi),
    \label{eq:epsilon_interp_main}
\end{equation}
where
\begin{equation}
    \bar{\epsilon}_i
    =
    \frac{\epsilon_i^s+\epsilon_i^l}{2},
    \qquad
    \Delta\epsilon_i
    =
    \epsilon_i^s-\epsilon_i^l.
    \label{eq:epsilon_definitions_main}
\end{equation}
The quantities \(\epsilon_i^s\) and \(\epsilon_i^l\) are the dilute solute free-energy offsets in the solid and liquid phases, respectively. The interpolation function is chosen as
\begin{equation}
    g(\phi)
    =
    \frac{15}{8}
    \left(
        \phi
        -
        \frac{2}{3}\phi^3
        +
        \frac{1}{5}\phi^5
    \right).
    \label{eq:g_function}
\end{equation}
This function satisfies \(g(\pm1)=\pm1\) and \(g'(\pm1)=g''(\pm1)=0\), which preserves the bulk free-energy minima in the solid and liquid phases.

The local interpolated grand-potential density of the dilute multi-component phase field model is given by
\begin{equation}
    \omega(\phi,\boldsymbol{\mu},T)
    =
    h f_{\rm dw}(\phi)
    +
    f_T(\phi)(T-T_m)
    -
    \sum_{i=1}^{n_s}
    \mathcal A_i
    \exp\left[
        \frac{\mu_i-\epsilon_i(\phi)}{\mathcal A_i}
    \right],
    \label{eq:gp_density_main}
\end{equation}
where
\begin{equation}
    f_{\rm dw}(\phi)
    =
    -\frac{\phi^2}{2}
    +
    \frac{\phi^4}{4}.
\end{equation}
Here \(h\) is the double-well barrier height, and \(f_T(\phi)\) is a thermal interpolation function, introduced below. From Eq.~\eqref{eq:gp_density_main} various properties can be derived.

The concentration associated with \(\mu_i\) follows from Eq.~\eqref{eq:gp_density_main} as
\begin{equation}
    c_i(\phi,\mu_i)
    =
    -\frac{\partial \omega}{\partial \mu_i}
    =
    \exp\left[
        \frac{\mu_i-\epsilon_i(\phi)}{\mathcal A_i}
    \right].
    \label{eq:c_mu_main}
\end{equation}
Since this relation is algebraically invertible, the diffusion potential of solute $i$ can be written as
\begin{equation}
    \mu_i
    =
    \mathcal A_i\ln c_i+\epsilon_i(\phi).
    \label{eq:mu_c_main}
\end{equation}
This concentration-variable representation is specific to the diagonal dilute approximation. In a fully coupled multicomponent formulation, the relation between \(\boldsymbol{\mu}\) and \(\mathbf c\) would generally require inversion of the full susceptibility matrix.

The partition coefficient of concentration between bulk phases is obtained by equating the diffusion potentials across an equilibrium solid-liquid interface, which gives
\begin{equation}
    k_i
    =
    \frac{c_i^s}{c_i^l}
    =
    \exp\left(
        -\frac{\Delta\epsilon_i}{\mathcal A_i}
    \right),
    \label{eq:ki_from_delta_epsilon_main}
\end{equation}
and hence
\begin{equation}
    \Delta\epsilon_i
    =
    -\mathcal A_i\ln k_i.
    \label{eq:delta_epsilon_main}
\end{equation}
It is useful to define
\begin{equation}
    b_i
    =
    \frac{1}{2}\ln k_i,
    \label{eq:b_i_main}
\end{equation}
so that Eq.~\eqref{eq:epsilon_interp_main} can also be written as
\begin{equation}
    \epsilon_i(\phi)
    =
    \bar{\epsilon}_i
    -
    \mathcal A_i b_i g(\phi).
    \label{eq:epsilon_bi_main}
\end{equation}

At a stationary equilibrium interface, \(\mu_i\) is spatially uniform. Taking the reference liquid concentration \(c_i^{l,0}\) as the normalization, Eq.~\eqref{eq:c_mu_main} yields the equilibrium concentration profile,
\begin{equation}
    c_i^0(\phi)
    =
    c_i^{l,0}
    \exp\left[
        b_i(1+g(\phi))
    \right].
    \label{eq:ci0_main}
\end{equation}
Thus \(c_i^0(-1)=c_i^{l,0}\) in the liquid and \(c_i^0(+1)=k_i c_i^{l,0}\) in the solid as expected.

The bulk liquidus hypersurface follows from equality of the solid and liquid grand-potential densities. Using Eq.~\eqref{eq:phase_gp_diagonal_dilute}, the bulk grand-potential density of phase \(\alpha=s,l\) can be written as
\begin{equation}
    \omega^\alpha
    =
    \omega_0^\alpha(T)
    -
    \sum_{i=1}^{n_s}
    \mathcal A_i c_i^\alpha.
\end{equation}
Near \(T_m\), the pure-solvent contribution is linearized as
\begin{equation}
    \omega_0^s(T)-\omega_0^l(T)
    =
    \Delta f_T(T-T_m),
    \label{eq:delta_fT_definition_main}
\end{equation}
where \(\Delta f_T\) is the pure-solvent thermal free-energy coefficient. Imposing \(\omega^s=\omega^l\) and \(c_i^s=k_i c_i^l\) gives
\begin{equation}
    T-T_m
    =
    -
    \sum_{i=1}^{n_s}
    \frac{\mathcal A_i(1-k_i)}{\Delta f_T}
    c_i^l.
    \label{eq:liquidus_from_A_main}
\end{equation}
The resulting dilute liquidus hypersurface is
\begin{equation}
    T-T_m
    =
    -
    \sum_{i=1}^{n_s}
    m_i c_i^l,
    \label{eq:target_liquidus_main}
\end{equation}
where \(m_i\) is the positive magnitude of the equilibrium liquidus slope of solute \(i\). Matching Eqs.~\eqref{eq:liquidus_from_A_main} and \eqref{eq:target_liquidus_main} gives
\begin{equation}
    \mathcal A_i
    =
    \frac{m_i\Delta f_T}{1-k_i}.
    \label{eq:A_i_matching_main}
\end{equation}
Equivalently, at the reference liquid concentration,
\begin{equation}
    X_i^{l,0}
    =
    \frac{c_i^{l,0}}{\mathcal A_i}
    =
    \frac{c_i^{l,0}(1-k_i)}
         {m_i\Delta f_T}.
    \label{eq:reference_susceptibility_main}
\end{equation}
The algebra leading to Eq.~\eqref{eq:A_i_matching_main}, and the equivalent expression for the reference susceptibility \(X_i^{l,0}\), is given in Supplementary Section~S2.
Equations~\eqref{eq:A_i_matching_main} and \eqref{eq:reference_susceptibility_main} show explicitly how component-specific liquidus slopes are represented by component-specific diagonal susceptibilities.

The thermal interpolation \(f_T(\phi)\) is chosen so that the reference equilibrium interface has the pure-material phase field profile when the concentration profiles in Eq.~\eqref{eq:ci0_main} are established. Defining the reference equilibrium liquidus undercooling
\begin{equation}
    M_0
    =
    \sum_{i=1}^{n_s}
    m_i c_i^{l,0}
    \label{eq:M0_main}
\end{equation}
this condition gives
\begin{equation}
    f_T'(\phi)
    =
    -
    \frac{g'(\phi)}{M_0}
    \sum_{i=1}^{n_s}
    \mathcal A_i b_i c_i^{l,0}
    \exp\left[
        b_i(1+g(\phi))
    \right].
    \label{eq:fT_prime_main}
\end{equation}
With the matching condition in Eq.~\eqref{eq:A_i_matching_main}, the jump \(f_T(+1)-f_T(-1)\) is consistent with \(\Delta f_T\) in Eq.~\eqref{eq:delta_fT_definition_main}. Details of the pointwise cancellation condition used to construct \(f_T(\phi)\), and the verification of the jump \(f_T(+1)-f_T(-1)=\Delta f_T\), are shown in Supplementary Section~S2.

\subsection{Evolution equations and interface anisotropy}
\label{subsec:evolution_equations_main}

We first write the evolution equations for the isotropic model with interface width \(W\) and relaxation time \(\tau_0\). The order-parameter dynamics are obtained by taking the variational derivative of the grand-potential functional
\begin{equation}
    \Omega[\phi,\boldsymbol{\mu},T]
    =
    \int_\Omega
    \left[
        \frac{hW^2}{2}|\nabla\phi|^2
        +
        \omega(\phi,\boldsymbol{\mu},T)
    \right]dV,
    \label{eq:Omega_functional_main}
\end{equation}
where \(\omega\) is given by Eq.~\eqref{eq:gp_density_main}. Using Eq.~\eqref{eq:fT_prime_main}, the order parameter equation can be written in terms of the concentration variables as
\begin{align}
    \tau_0
    \frac{\partial\phi}{\partial t}
    &=
    W^2\nabla^2\phi
    +
    \phi-\phi^3 \nonumber\\
    &\qquad-
    g'(\phi)
    \sum_{i=1}^{n_s}
    \lambda_i
    \left[
        c_i
        +
        \frac{T-T_m}{M_0}
        c_i^{l,0}
        \exp\left[
            b_i(1+g(\phi))
        \right]
    \right],
    \label{eq:phi_isotropic_main}
\end{align}
where the component-specific coupling constants are
\begin{equation}
    \lambda_i
    =
    -\frac{a_1^0 W}{\Gamma}
    \frac{m_i b_i}{1-k_i},
    \qquad
    a_1^0=\frac{2\sqrt{2}}{3},
    \label{eq:lambda_i_main}
\end{equation}
and where \(\Gamma\) is the Gibbs-Thomson coefficient. For the usual case \(k_i<1\), one has \(b_i<0\), and Eq.~\eqref{eq:lambda_i_main} gives \(\lambda_i>0\). The time scale is matched to the reference kinetic coefficient \(\mu_0\) through
\begin{equation}
    \tau_0
    =
    \frac{W^2}{\Gamma\mu_0}.
    \label{eq:tau0_main}
\end{equation}
The one-dimensional solvability calculation leading to Eqs.~\eqref{eq:lambda_i_main} and \eqref{eq:tau0_main} is given in Supplementary Section~S3.

The concentration equations are written in conserved form. Using the diffusion potential in Eq.~\eqref{eq:mu_c_main} and choosing the mobility
\begin{equation}
    K_i
    =
    \frac{D_l^i q_i(\phi)c_i}{\mathcal A_i},
\end{equation}
where \(D_l^i\) is the liquid diffusivity of solute \(i\), gives
\begin{equation}
    \frac{\partial c_i}{\partial t}
    =
    \nabla\cdot
    \left\{
        D_l^i q_i(\phi)c_i
        \nabla
        \left[
            \ln c_i-b_i g(\phi)
        \right]
    \right\},
    \qquad
    i=1,\ldots,n_s .
    \label{eq:ci_multicomponent_main}
\end{equation}
The functions \(q_i(\phi)\) interpolate the solute mobility through the diffuse interface. For a one-sided model with negligible solid diffusivity, they satisfy
\begin{equation}
    q_i(-1)=1,
    \qquad
    q_i(+1)=0.
    \label{eq:q_boundary_conditions_main}
\end{equation}
A commonly used one-sided interpolation satisfying this criterion is a simple linear function:
\begin{equation}
    q_{\rm lin}(\phi)
    =
    \frac{1-\phi}{2}.
    \label{eq:q_linear_main}
\end{equation}
In the present work, \(q_i(\phi)\) is not fixed to this form. Instead, it is optimized so that the one-dimensional phase field response reproduces the prescribed nonequilibrium functions \(k_i(V)\) and \(m_i(V)\), as described in Sec.~\ref{sec:q_optimization}.

For multidimensional simulations, surface-energy and kinetic anisotropy are included by replacing the isotropic width and relaxation time in Eq.~\eqref{eq:phi_isotropic_main} by orientation-dependent quantities. The interface width is written as
\begin{equation}
    W(\mathbf n)
    =
    S W_0 a_s(\mathbf n),
    \label{eq:W_anisotropic_main}
\end{equation}
where \(W_0\) is the physical interface-width scale, \(S\) is the upscaling factor, and \(a_s(\mathbf n)\) describes the surface-energy anisotropy. The relaxation time is written as
\begin{equation}
    \tau(\mathbf n)
    =
    \tau_0
    \frac{a_s(\mathbf n)^2}{a_k(\mathbf n)},
    \qquad
    \tau_0
    =
    \frac{(S W_0)^2}{\Gamma\mu_0},
    \label{eq:tau_anisotropic_main}
\end{equation}
where \(a_k(\mathbf n)\) describes the kinetic anisotropy and $\mathbf n =-\nabla \phi /|\nabla \phi|$ defines the unit vector normal to the interface. The phase field equation used in two-dimensional anisotropic simulations thus becomes
\begin{align}
    \tau(\mathbf n)
    \frac{\partial\phi}{\partial t}
    &=
    \nabla\cdot
    \left[
        W(\mathbf n)^2\nabla\phi
    \right]
    +
    \phi-\phi^3
    +
    \sum_{j=x,y}
    \partial_j
    \left[
        |\nabla\phi|^2
        W(\mathbf n)
        \frac{\partial W(\mathbf n)}
        {\partial(\partial_j\phi)}
    \right]
    \nonumber\\
    &\qquad
    -
    g'(\phi)
    \sum_{i=1}^{n_s}
    \lambda_i
    \left[
        c_i
        +
        \frac{T-T_m}{M_0}
        c_i^{l,0}
        \exp\left[
            b_i(1+g(\phi))
        \right]
    \right].
    \label{eq:phi_anisotropic_main}
\end{align}
The anisotropy functions are taken to be fourfold,
\begin{align}
    a_s(\mathbf n)
    &=
    1+\epsilon_s\cos(4\theta),
    \label{eq:as_main}
    \\
    a_k(\mathbf n)
    &=
    1+\epsilon_k\cos(4\theta),
    \label{eq:ak_main}
\end{align}
where \(\epsilon_s\) and \(\epsilon_k\) are the surface-energy and kinetic anisotropy strengths. The angle \(\theta\) is measured between the interface normal and the \(x\)-axis. In terms of order parameter gradients, it may be evaluated as
\begin{equation}
    \theta
    =
    \tan^{-1}
    \left(
        \frac{\partial_y\phi}{\partial_x\phi}
    \right),
    \label{eq:theta_main}
\end{equation}
up to the convention chosen for the sign of the interface normal. In Eq.~\eqref{eq:phi_anisotropic_main}, the coupling constants \(\lambda_i\) are evaluated using the reference width \(W=SW_0\), while anisotropy enters through the orientation dependence of \(W(\mathbf n)\) and \(\tau(\mathbf n)\).

\subsection{Planar-interface response functions}
\label{subsec:response_functions_main}

The optimized diffusivity functions \(q_i(\phi)\) are calibrated using one-dimensional steady planar-interface solutions. For a moving interface, the phase field prediction of the partition coefficient is defined by
\begin{equation}
    k_i^{\rm PF}(V)
    =
    \frac{c_i^s}{c_i^l},
    \label{eq:ki_pf_main}
\end{equation}
where \(c_i^s\) is the solid-side concentration incorporated into the solid, and \(c_i^l\) is the liquid-side concentration at the interface. In the one-sided steady calibration problem, \(c_i^s\) is fixed by mass conservation and equals the imposed nominal composition \(c_{\infty,i}\).

The phase field prediction of the kinetic liquidus slope is obtained from the one-dimensional steady-state solvability condition. Multiplying the moving-frame form of Eq.~\eqref{eq:phi_isotropic_main} by \(d\phi/dx\) and integrating through the interface gives
\begin{equation}
    \frac{m_i^{\rm PF}(V)}{m_i}
    =
    \frac{b_i}{(1-k_i)c_i^l}
    \int_{-\infty}^{+\infty}
    g'(\phi)c_i
    \frac{d\phi}{dx}\,dx .
    \label{eq:mi_pf_main}
\end{equation}
The derivation of Eq.~\eqref{eq:mi_pf_main} from the compositional term in the complete Gibbs-Thomson conditions is given in Supplementary Section~S3.
For an interface in equilibrium, Eq.~\eqref{eq:mi_pf_main} reduces to \(m_i^{\rm PF}(0)=m_i\) for each solute. Equations~\eqref{eq:ki_pf_main} and \eqref{eq:mi_pf_main} provide the bridge between the multicomponent phase field model and the sharp-interface response functions such as the CGM model. The functions \(q_i(\phi)\) determine the concentration profiles through the moving diffuse interface and therefore control both \(k_i^{\rm PF}(V)\) and \(m_i^{\rm PF}(V)\). 
For a single dilute solute, the sums in the model equations contain one term, and the formulation reduces to the dilute binary far-from-equilibrium phase field model with enhanced interfacial solute diffusivity as derived in Refs.~\cite{Ji2023,Ji2025}.

The next section formulates the selection of \(q_i(\phi)\) as an optimization problem in which these phase field response functions are matched to prescribed targets.

\section{Optimization of interfacial diffusivity functions}
\label{sec:q_optimization}

The multicomponent phase field formulation in Sec.~\ref{sec:model_development} contains one interfacial diffusivity interpolation \(q_i(\phi)\) for each dilute solute species. These functions determine the effective solute redistribution kinetics inside the diffuse solid-liquid interface and therefore control the phase field predictions of the velocity-dependent partition coefficients and kinetic liquidus slopes. In a one-sided model with negligible solid diffusivity, each interpolation must satisfy \(q_i(-1)=1\) and \(q_i(+1)=0\) so that the diffusivity becomes \(D_l^i\) in the bulk liquid, and vanishes in the solid. The simplest choice is the linear interpolation in Eq.~\eqref{eq:q_linear_main}. However, when the phase field interface thickness is increased from the physical value \(W_0\) to a computationally tractable value \(W=S W_0\), the effective time available for solute redistribution across the diffuse interface is changed. If \(q_i(\phi)\) is left unchanged, this produces excess solute trapping relative to the desired sharp-interface response, i.e., \(k_i(V)\). Enhanced interfacial diffusivity functions can compensate this effect by increasing solute mobility within the diffuse interface while leaving the bulk diffusivities unchanged~\cite{Ji2023}.

In this section, the objective is to find a smooth endpoint-preserving diffusivity profile  \(q_i(\phi)\) such that one-dimensional steady planar-interface solutions reproduce prescribed target response functions, \(k_{i,\rm tar}(V)\) and \(m_{i,\rm tar}(V)\) for each solute species. In the examples considered below, these targets are generated from the continuous growth model response functions introduced in Sec.~\ref{sec:introduction}: Eqs.~\eqref{eq:intro_cgm_ki} and~\eqref{eq:intro_cgm_mi}. More generally, the same procedure can use target functions obtained from sharp-interface theory, atomistic simulations, experiments, or reference phase field calculations performed at the physical interface width.

\subsection{Endpoint-preserving representation}
\label{subsec:q_representation}

For each solute, the optimized diffusivity interpolation is written as a correction to the linear one-sided profile,
\begin{equation}
    q_{i,N}(\phi)
    =
    \frac{1-\phi}{2}
    +
    (1-\phi^2) p_{i,N}(\phi),
    \label{eq:q_additive_linear}
\end{equation}
where
\begin{equation}
    p_{i,N}(\phi)
    =
    \sum_{n=0}^{N-1} a_{i,n} T_n(\phi).
    \label{eq:chebyshev_expansion}
\end{equation}
Here \(T_n(\phi)\) is the Chebyshev polynomial of degree \(n\), \(a_{i,n}\) are optimization coefficients, and \(N\) is the number of basis functions. Chebyshev polynomials are used because the order parameter, by definition, lies in the same interval of \(\phi\in[-1,+1]\) while providing a flexible representation of the interface diffusivity. The specific functional form used, however, is not essential for the presented optimization framework. The prefactor \(1-\phi^2\) ensures that the correction vanishes in the two bulk phases. Consequently, the endpoint conditions in Eq.~\eqref{eq:q_boundary_conditions_main} are satisfied exactly for any values of the coefficients \(a_{i,n}\). The optimization therefore changes only the interfacial part of the solute diffusivity profile.

This additive-linear representation provides a controlled generalization of the standard one-sided diffusion interpolation. The first term in Eq.~\eqref{eq:q_additive_linear} fixes the bulk liquid and solid limits, while the second term supplies the additional interfacial mobility required to reproduce the desired nonequilibrium response. Increasing \(N\) increases the flexibility of the interfacial correction in a systematic way. In practice, \(N\) is chosen large enough to represent the target response functions accurately, while a smoothness regularization term discussed below is used to avoid unnecessarily sharp changes in \(q_i(\phi)\).

\subsection{One-dimensional response-function calibration}
\label{subsec:target_functions}

For a trial set of coefficients
\begin{equation}
    \mathbf a_i
    =
    (a_{i,0},a_{i,1},\ldots,a_{i,N-1}),
\end{equation}
the corresponding phase field response functions are computed from a one-dimensional steady planar-interface calculation. For each velocity \(V\), the calculation gives \(k_i^{\rm PF}(V;\mathbf a_i)\) and \(m_i^{\rm PF}(V;\mathbf a_i)\) which are compared directly with the prescribed target functions \(k_{i,\rm tar}(V)\) and \(m_{i,\rm tar}(V)\). These are intrinsic planar-interface response functions and can therefore be calibrated in one dimension before the optimized diffusivity interpolations are used in multidimensional simulations.

The calibration is formulated in the dimensionless moving-frame coordinate \(\xi={x}/{W}\) with \(W=S W_0\) and approximates the stationary interface profile as
\begin{equation}
    \phi_0(\xi)
    =
    -\tanh\left(
        \frac{\xi}{\sqrt{2}}
    \right).
    \label{eq:stationary_phi_xi}
\end{equation}
The solid and liquid correspond to \(\xi\to-\infty\) and \(\xi\to+\infty\), respectively. For solute \(i\), the steady-state concentration equation in the moving frame of the interface is written as
\begin{equation}
    -\frac{S W_0 V}{D_l^i}
    \frac{dc_i}{d\xi}
    =
    \frac{d}{d\xi}
    \left[
        q_{i,N}(\phi_0)c_i
        \frac{d}{d\xi}
        \left(
            \ln c_i
            -
            b_i g(\phi_0)
        \right)
    \right].
    \label{eq:steady_concentration_conservation_opt}
\end{equation}
Equation~\eqref{eq:steady_concentration_conservation_opt} is the nondimensional form of the moving-frame dilute diffusion equation. The far-solid condition is \(c_i(\xi\to-\infty)=c_{\infty,i}\), and the integration constant is chosen so that the diffusive flux vanishes in the far liquid.

After one integration, Eq.~\eqref{eq:steady_concentration_conservation_opt} can be transformed into a first-order equation,
\begin{equation}
    \frac{dc_i}{d\xi}
    =
    b_i c_i g'(\phi_0)
    \frac{d\phi_0}{d\xi}
    -
    \frac{SW_0 V}{D_l^iq_{i,N}(\phi_0)}
    \left(
        c_i-c_{\infty,i}
    \right).
    \label{eq:dc_dxi_opt}
\end{equation}
The first term in Eq.~\eqref{eq:dc_dxi_opt} drives the concentration toward the equilibrium diffuse-interface partitioning profile, while the second term represents the advective departure from that profile caused by interface motion. Increasing \(q_i(\phi)\) inside the interface increases the effective redistribution rate and thereby compensates the excess trapping introduced by the enlarged interface width.

The phase field partition coefficient is extracted from the liquid-side interface concentration obtained from the steady profile,
\begin{equation}
    k_i^{\rm PF}(V;\mathbf a_i)
    =
    \frac{c_{\infty,i}}{c_i^l(V;\mathbf a_i)},
    \label{eq:k_pf_from_profile_opt}
\end{equation}
where, in a one-sided steady-state problem, mass conservation fixes the concentration incorporated into the solid to the imposed nominal value, \(c_i^s=c_{\infty,i}\). The liquid-side concentration \(c_i^l\) is then taken as the solute peak at the interface, which corresponds to the maximum of the steady concentration profile. The kinetic liquidus-slope response is computed from the corresponding integrated thermodynamic contribution given by Eq.~\eqref{eq:mi_pf_main} in dimensionless form:
\begin{equation}
    \frac{
    m_i^{\rm PF}(V;\mathbf a_i)
    }{
    m_i
    }
    =
    \frac{
    b_i
    }{
    (1-k_i)c_i^l(V;\mathbf a_i)
    }
    \int_{-\infty}^{+\infty}
    g'(\phi_0)c_i(\xi;V,\mathbf a_i)
    \frac{d\phi_0}{d\xi}
    \,d\xi .
    \label{eq:m_pf_from_profile_opt}
\end{equation}
Equations~\eqref{eq:k_pf_from_profile_opt} and \eqref{eq:m_pf_from_profile_opt} define the phase field response functions used in the objective function below. The use of the fixed profile \(\phi_0\) is an approximation to the full moving-interface phase field solution, but it yields an efficient and robust calibration problem for determining interfacial diffusivity profiles.

In the dilute independent-solute limit, this calibration is diagonal in solute index. The same one-dimensional inverse problem is therefore solved separately for each \(q_i(\phi)\), using the corresponding \(k_i\), \(m_i\), \(V_{D,i}\), \(k_{i,\rm tar}(V)\), and \(m_{i,\rm tar}(V)\). The optimized functions are then assembled and used in the multicomponent phase field equations derived in Sec.~\ref{sec:model_development}.

\subsection{Loss function and smoothness regularization}
\label{subsec:objective_function}

The coefficients are optimized by minimizing a loss function that balances response-function accuracy against smoothness of the interfacial diffusivity interpolation. Since the dilute independent-solute approximation diagonalizes the calibration problem, the loss can be minimized independently for each solute. For one solute, the data-misfit part of the loss is
\begin{align}
    \mathcal L_{km}^{(i)}(\mathbf a_i)
    &=
    \left\langle
    \left[
    \frac{
    k_i^{\rm PF}(V_j;\mathbf a_i)-k_{i,\rm tar}(V_j)
    }{
    k_{i,\rm tar}(V_j)
    }
    \right]^2
    \right\rangle_j \nonumber\\
    &\qquad +
    w_m
    \left\langle
    \left[
    \frac{
    m_i^{\rm PF}(V_j;\mathbf a_i)-m_{i,\rm tar}(V_j)
    }{
    m_{i,\rm tar}(V_j)
    }
    \right]^2
    \right\rangle_j .
    \label{eq:data_loss_single}
\end{align}
Here \(V_j\) are the calibration velocities, \(\langle\cdot\rangle_j\) denotes an average over those velocities, and \(w_m\) controls the relative weight assigned to the kinetic liquidus-slope error. Unless otherwise stated, \(w_m=1\), giving equal weights.

Smoothness is controlled by penalizing the curvature of the diffusivity interpolation,
\begin{equation}
    \mathcal R_{{\rm curv},i}
    =
    \left\langle
    \left[
    \frac{d^2 q_{i,N}}{d\phi^2}
    \right]^2
    \right\rangle_\phi ,
    \label{eq:curvature_penalty_single}
\end{equation}
where the average is evaluated over \(\phi\in[-1,1]\). The regularized single-solute loss is
\begin{equation}
    \mathcal L_i
    =
    \mathcal L_{km}^{(i)}
    +
    \lambda_{{\rm curv},i}\mathcal R_{{\rm curv},i}.
    \label{eq:total_loss_single}
\end{equation}
The regularization parameter \(\lambda_{{\rm curv},i}\) controls the tradeoff between matching the target response and suppressing sharp variations of $q_{i,N}(\phi)$. Small values of \(\lambda_{{\rm curv},i}\) prioritize accuracy in \(k_i(V)\) and \(m_i(V)\), whereas larger values produce smoother diffusivity profiles. This tradeoff is important in multidimensional simulations because an unnecessarily sharp or high-amplitude interfacial diffusivity enhancement can be difficult to resolve and may introduce unwanted transport along the diffuse interface.

For a multicomponent calculation, Eq.~\eqref{eq:total_loss_single} is minimized separately for each solute in the dilute independent-solute approximation. If desired, the individual losses may be summed to define a combined diagnostic,
\begin{equation}
    \mathcal L
    =
    \sum_{i=1}^{n_s}
    \mathcal L_i,
    \label{eq:total_loss_multicomponent_diagnostic}
\end{equation}
but no coupled optimization is required in the present dilute formulation.

\subsection{Tolerance-based selection of the optimized interpolation}
\label{subsec:tolerance_selection}

Rather than selecting \(\lambda_{{\rm curv},i}\) manually, we use a tolerance-based selection rule. A set of candidate regularization strengths,
\begin{equation}
    \lambda_{{\rm curv},i}^{(1)},
    \lambda_{{\rm curv},i}^{(2)},
    \ldots,
    \lambda_{{\rm curv},i}^{(M)},
\end{equation}
is considered. For each regularization value, Eq.~\eqref{eq:total_loss_single} is minimized, and the resulting interpolation \(q_{i,N}(\phi)\) and discrete response values \(k_i^{\rm PF}(V_j)\) and \(m_i^{\rm PF}(V_j)\) are recorded on the calibration velocity grid. Candidate solutions are then filtered according to prescribed accuracy tolerances. For each solute, we define the root-mean-squared errors
\begin{align}
    \mathrm{RMS}_{k,i}
    &=
    \left[
    \left\langle
    \left(
    \frac{
    k_i^{\rm PF}(V_j)-k_{i,\rm tar}(V_j)
    }{
    k_{i,\rm tar}(V_j)
    }
    \right)^2
    \right\rangle_j
    \right]^{1/2},
    \label{eq:rms_k_i}
    \\
    \mathrm{RMS}_{m,i}
    &=
    \left[
    \left\langle
    \left(
    \frac{
    m_i^{\rm PF}(V_j)-m_{i,\rm tar}(V_j)
    }{
    m_{i,\rm tar}(V_j)
    }
    \right)^2
    \right\rangle_j
    \right]^{1/2}.
    \label{eq:rms_m_i}
\end{align}
A candidate interpolation is accepted if
\begin{equation}
    \mathrm{RMS}_{k,i}\leq\varepsilon_{\rm RMS},
    \qquad
    \mathrm{RMS}_{m,i}\leq\varepsilon_{\rm RMS}.
    \label{eq:rms_acceptance_opt}
\end{equation}

The RMS tolerance is the primary accuracy criterion used in this work. In applications where a particular velocity interval is more important than the full calibration range, the same framework can be applied with nonuniform sampling of the velocities \(V_j\), velocity-dependent weights in the averages entering Eq.~\eqref{eq:data_loss_single}, or by restricting the optimization to the velocity range of interest. Similarly, if required, the accepted candidates can be further screened by imposing a maximum relative error tolerance on \(k_i^{\rm PF}(V_j)\) and \(m_i^{\rm PF}(V_j)\) over the sampled velocities. These additional choices do not change the structure of the optimization problem; they only specify which part of the response curve is prioritized for a given application.

Among the accepted candidates, we choose the interpolation with the smallest curvature penalty \(\mathcal R_{{\rm curv},i}\). The selected function is therefore the smoothest \(q_i(\phi)\) that reproduces the prescribed response functions within the specified accuracy. In this sense, the tolerance \(\varepsilon_{\rm RMS}\) acts as a response-function accuracy criterion, while the regularization sweep determines the smoothest diffuse-interface realization of that accuracy. This selection rule avoids choosing an unnecessarily sharp interfacial diffusivity enhancement simply because it gives a marginally smaller one-dimensional fitting error.

The same procedure applies to binary and multicomponent dilute systems. In the binary case, a single interpolation \(q(\phi)\) is optimized to match \(k(V)\) and \(m(V)\). In the multicomponent case, separate interpolations \(q_i(\phi)\) are optimized for the individual solute response functions and then used together in the multicomponent phase field equations. This provides a direct route for embedding prescribed solute-specific nonequilibrium interface kinetics into multidimensional simulations.

The additive representation in Eq.~\eqref{eq:q_additive_linear} is used throughout this work because it is simple, endpoint preserving, and sufficiently flexible for the response functions considered here. Since this representation does not enforce positivity for arbitrary coefficient values, we also tested a positivity-preserving logarithmic representation,
\begin{equation}
    q_{i,N}^{\rm log}(\phi)
    =
    \frac{1-\phi}{2}
    \exp\left[
        \frac{1+\phi}{2}
        p_{i,N}(\phi)
    \right],
    \label{eq:q_exp_log}
\end{equation}
with the same Chebyshev expansion \(p_{i,N}(\phi)\) as in Eq.~\eqref{eq:chebyshev_expansion}. This form satisfies \(q_{i,N}^{\rm log}(-1)=1\), \(q_{i,N}^{\rm log}(+1)=0\), and remains positive for \(-1<\phi<1\). For the cases examined here, the logarithmic representation produced interpolation functions and response-function fits similar to those obtained with the additive form. We additionally tested more flexible neural-network representations of \(q_i(\phi)\), utilizing multilayer-perceptron parameterizations, and obtained qualitatively similar optimized diffusivity profiles. This similarity suggests that simultaneous matching of \(k_i(V)\) and \(m_i(V)\) over the calibration range strongly constrains the admissible interfacial diffusivity profile, making the optimized result only weakly dependent on the chosen representation for the cases studied here. We use the additive Chebyshev representation of Eq.~\eqref{eq:q_additive_linear} in the remainder of this work because it provides the same practical response-function matching while remaining transparent, compact, and easy to regularize.

\section{Results}
\label{sec:results}

\subsection{Response-function matching in binary Al--Cu}
\label{subsec:binary_alcu_matching}

We first apply the above calibration procedure to a dilute binary Al--Cu alloy. This case provides a controlled setting for testing whether the optimized interfacial diffusivity interpolation can reproduce prescribed velocity-dependent partitioning and drag-modified kinetic liquidus responses before the resulting functions are used in multidimensional simulations. The target response functions are taken from the CGM expressions introduced above, using the material parameters used in Ref.~\cite{Ji2023}, listed in Table~\ref{tab:alcu_binary_parameters}. The computational interface thickness is \(W=S W_0\) with \(S=5\). The diffusive speed entering the target CGM response is taken as \(V_D=2.3~\mathrm{m\,s^{-1}}\) from atomistic simulations carried out in Ref.~\cite{Haapalehto2022}, close to the commonly used approximation \(V_D\sim D_l/W_0=2.4~\mathrm{m\,s^{-1}}\), assuming $W_0 \sim 1~\mathrm{nm}$.

\begin{table}[ht!]\small
    \centering
    \caption{
    Material parameters used for the binary Al--Cu alloy.
    }
    \label{tab:alcu_binary_parameters}
    \begin{tabular}{lll}
        \hline
        Symbol & Value & Description  \\
        \hline
        \(k_\mathrm{Cu}\) & \(0.14\) & Equilibrium partition coefficient \\
        \(m_\mathrm{Cu}\) & \(2.6~\mathrm{K\,wt.\%^{-1}}\) & Equilibrium liquidus slope  \\
        \(T_m\) & \(933~\mathrm{K}\) & Pure Al melting point  \\
        \(D_{l,\mathrm{Cu}}\) & \(2.4\times10^{-9}~\mathrm{m^2\,s^{-1}}\) & Diffusion in liquid \\
        \(V_{D,\mathrm{Cu}}\) & \(2.3~\mathrm{m\,s^{-1}}\) & Diffusive velocity \\
        \(\Gamma\) & \(2.4\times10^{-7}~\mathrm{K\,m}\) & Gibbs-Thomson coefficient \\
        \(\epsilon_s\) & \(0.012\) & Surface energy anisotropy \\
        \(\mu_0\) & \(0.5~\mathrm{m\,s^{-1}\,K^{-1}}\) & Kinetic coefficient \\
        \(\epsilon_k\) & \(0.10\) & Kinetic anisotropy \\
        \(c_\infty\) & \(3~\mathrm{wt.\%}\) & Nominal concentration  \\
        \hline
    \end{tabular}
\end{table}

For each drag coefficient \(\alpha\), the optimized function \(q(\phi)\) was obtained from the one-dimensional steady planar-interface calibration described in Sec.~\ref{subsec:target_functions}. In the present calculations, \(q(\phi)\) was represented using the additive-linear Chebyshev form of Eq.~\eqref{eq:q_additive_linear} with \(N=8\) basis functions, which provided sufficient flexibility while maintaining an efficient optimization. The contributions of the individual terms and the cumulative representation are shown in Supplementary Section~S4. The endpoint behavior was fixed so that the bulk diffusivities were unchanged, and only the interfacial diffusivity interpolation was modified. The calibration velocities were logarithmically spaced over \(0.01\le V\le 1.0~\mathrm{m\,s^{-1}}\) using \(50\) velocity points. This range covers the velocities relevant to the directional-solidification simulations considered below, for which the imposed thermal-gradient pulling speed is \(V_\mathrm{pull}=\)0.08--0.20~\(\mathrm{m\,s^{-1}}\). The logarithmic sampling gives approximately equal weight to each decade in velocity and therefore places a denser set of points at low velocities when viewed on a linear velocity scale. This is useful because the transition away from local equilibrium begins at low and intermediate velocities, where the relative changes in both \(k(V)\) and \(m(V;\alpha)\) are most pronounced.

Equal weight was assigned to the relative errors in \(k(V)\) and \(m(V;\alpha)\). The target tolerance used for the production interpolations was \(1\%\), which was found to provide a useful compromise between response-function accuracy and smoothness of \(q(\phi)\). Lower tolerances force the optimizer to reproduce the target curves more tightly but can introduce sharper features in the diffusivity interpolation profile, while looser tolerances produce smoother interpolations at the expense of reduced response-function accuracy. As a check on the fixed $\phi$-profile approximation used in the calibration, we repeated representative optimizations using numerically relaxed velocity-dependent phase field profiles instead of the analytic profile in Eq.~\eqref{eq:stationary_phi_xi}. The selected regularization, optimized \(q(\phi)\), and response-function errors were essentially unchanged, so the analytic profile was used in the optimizations for computational efficiency.

\begin{figure}[t]
    \centering
    \includegraphics[]{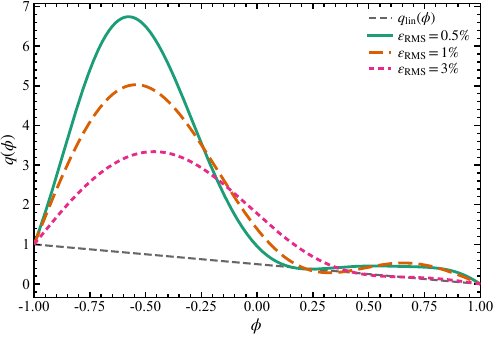}
    \caption{
    Effect of the response-function tolerance on the optimized interfacial diffusivity for the binary Al--Cu calibration with \(\alpha=0.3\). The optimized \(q(\phi)\) profiles are compared with the linear interpolation \(q_{\rm lin}(\phi)\).
    }
    \label{fig:alcu_tolerance_effect}
\end{figure}

\begin{figure*}[ht!]
    \centering
    \includegraphics[]{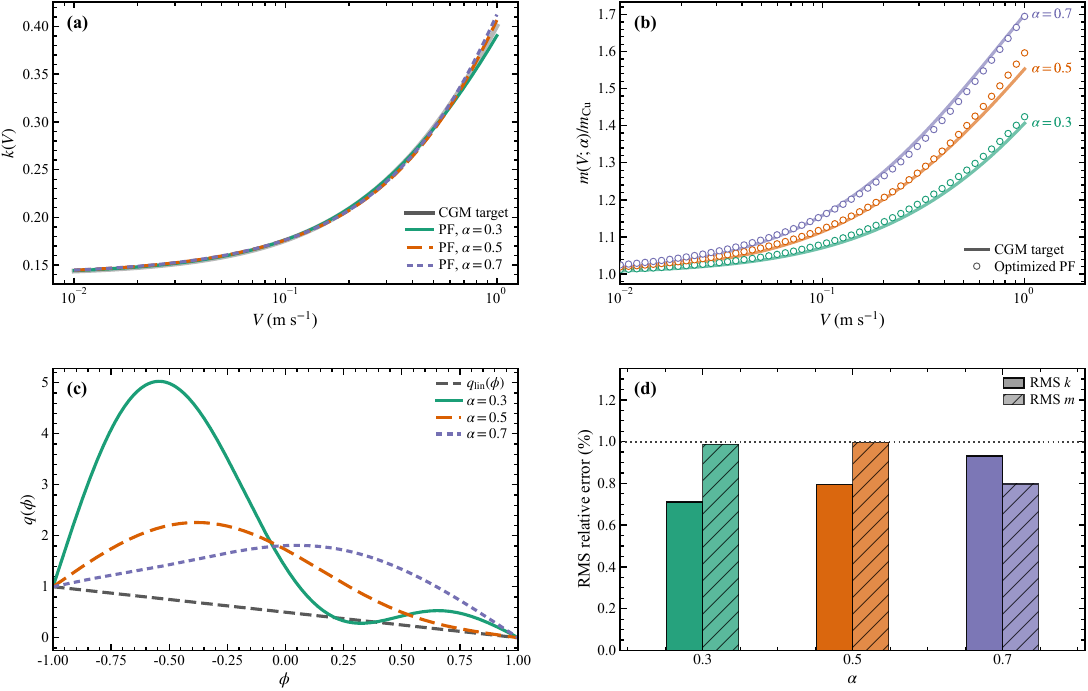}
    \caption{
    Binary Al--Cu response-function matching for optimized additive-linear Chebyshev interpolations with \(N=8\).
    (a) CGM target partition coefficient \(k_{\rm tar}(V)\) and optimized phase field (PF) values \(k^{\rm PF}(V)\) with varying values of solute drag parameter $\alpha$, where the target \(k_{\rm tar}(V)\) is common to all drag coefficients.
    (b) Drag-dependent target kinetic liquidus response \(m_{\rm tar}(V;\alpha)/m_{\rm Cu}\) and optimized phase field values \(m^{\rm PF}(V;\alpha)/m_{\rm Cu}\).
    Solid lines denote CGM targets and symbols denote phase field values extracted from the one-dimensional steady profiles. 
    (c) Optimized interfacial diffusivity functions \(q(\phi)\), compared with the linear interpolation \(q_{\rm lin}(\phi)\).
    (d) Root-mean-square (RMS) relative errors in \(k(V)\) and \(m(V;\alpha)\) for the optimized interpolations, where
    the dashed line indicates the selected \(1\%\) tolerance limit.
    }
    \label{fig:alcu_response_matching}
\end{figure*}

Figure~\ref{fig:alcu_tolerance_effect} illustrates the tradeoff of using different tolerances for the \(\alpha=0.3\) drag case. This case was chosen because it is one of the values used below in the two-dimensional simulations and provides a clear example of how the tolerance affects the optimized interpolation. Tightening the tolerance modifies the shape of \(q(\phi)\) in the diffuse-interface region, where the enhanced diffusivity compensates for the enlarged interface width. 
The strictest tolerance produces a profile with more pronounced local variations in \(q(\phi)\), whereas the looser tolerance permits a smoother profile at the expense of a larger deviation from the target response functions. Since the optimized \(q(\phi)\) is later tabulated and used directly in multidimensional simulations, excessive structure in the interpolation is undesirable as it can increase sensitivity to spatial resolution and may require smaller grid spacings to resolve the intended interpolation profile. We therefore use the \(1\%\) tolerance in the simulations below.
The optimized response functions for the drag coefficients used in the binary simulations discussed below are summarized in Fig.~\ref{fig:alcu_response_matching}. The target partition coefficient \(k_{\rm tar}(V)\) is independent of \(\alpha\), and the optimized phase field values collapse onto the same CGM curve for all drag coefficients with good agreement. In contrast, the target \(m_{\rm tar}(V;\alpha)\) depends strongly on \(\alpha\). Increasing \(\alpha\) increases the kinetic liquidus response over the partial-trapping regime, reflecting the larger fraction of the redistribution contribution subtracted from the driving force for interface motion. The optimized phase field values reproduce this systematic drag dependence over the full calibration range while satisfying the imposed common partitioning response. This controlled construction allows the multidimensional simulations below to isolate the consequences of the drag-modified kinetic liquidus response from changes caused by different solute-trapping kinetics.
The corresponding \(q(\phi)\) profiles differ within the diffuse-interface region, retain the prescribed endpoint behavior in the bulk phases, and remain positive over \(-1\leq\phi\leq+1\). This is the desired outcome of the calibration: the optimized interpolation modifies the effective redistribution kinetics only where the enlarged phase field interface would otherwise produce excess trapping. Fig.~\ref{fig:alcu_response_matching}(d) shows that both response functions are reproduced within the prescribed tolerance computed from Eqs.~\eqref{eq:rms_k_i} and \eqref{eq:rms_m_i} for the drag coefficients considered here. The optimized functions are therefore sufficiently accurate for testing how the prescribed drag-modified response influences two-dimensional microstructure selection.

The attainable response range depends on the chosen calibration degrees of freedom. In the present implementation and alloy used, the additive-linear \(q(\phi)\) representation with fixed thermodynamic interpolation \(g(\phi)\) fails to capture both \(k_{\rm tar}(V)\) and \(m_{\rm tar}(V;\alpha)\) for very small and very large values of \(\alpha\). In those regimes, the optimizer tends either to require sharper interfacial diffusivity features near \(\phi=\pm1\) or to encounter response-function combinations that are difficult to reproduce using \(q(\phi)\) alone. This limitation is not unexpected: the present formulation uses the diffusivity interpolation as the only calibration degree of freedom, while the thermodynamic interpolation remains fixed. Allowing \(g(\phi)\) to vary together with \(q(\phi)\), or introducing a more flexible constrained representation, may enlarge the attainable range of response functions. We leave this extension for future work.

For this reason, the binary demonstrations below focus on the intermediate range \(0.3\le \alpha\le0.8\). This interval is also physically expected for the present purpose: partial solute drag is expected to lie between the limiting no-drag and full-drag cases, and prior atomistic and theoretical studies motivate intermediate values in this range for binary alloys rather than either extreme~\cite{Haapalehto2022,Antillon2023,Ahmad1998,Ji2023}. The selected range therefore provides a controlled setting for the demonstration below: whether quantitative differences in the prescribed kinetic liquidus response, at a fixed velocity-dependent partition coefficient, propagate into the morphologies predicted by multidimensional phase field simulations.

As a robustness check for the following two-dimensional simulations, the optimized interpolations were also evaluated without refitting beyond the calibration interval. Over the extrapolated range up to \(10~\mathrm{m\,s^{-1}}\), the extrapolated responses remained smooth and CGM-like, with typical RMS relative errors increasing only to a few percent for both \(k(V)\) and \(m(V;\alpha)\), and hence having substantially smaller deviations than the corresponding linear-\(q(\phi)\) baseline with significant excess trapping.

\subsection{Drag-dependent two-dimensional Al--Cu microstructures}

The optimized binary Al--Cu response functions were next used in two-dimensional directional-solidification simulations to test how the imposed solute drag response affects the growth morphologies. The purpose of these simulations is not to construct a complete processing map, but rather to isolate the effect of changing \(m(V;\alpha)\) while keeping the same target partition coefficient \(k(V)\). All simulations in this subsection therefore use the same alloy parameters, nominal composition, thermal gradient, and numerical domain, while only the pulling velocity and the optimized \(q(\phi)\) corresponding to the chosen drag coefficient is changed.

The simulations were carried out in a moving frame using a rectangular domain of size \(15\times 30~\mu\mathrm{m}^2\). The thermal gradient was set to \(G=5\times10^6~\mathrm{K\,m^{-1}}\), representing typical AM conditions~\cite{nanotoday,Ji2024_banding,Ji2025}.
The temperature evolution follows the {frozen temperature approximation}, written as
\begin{equation}\label{eq:FTA}
    T = T_0 + G(x-x_0-V_\mathrm{pull}t),
\end{equation}
where $T_0$ is a reference temperature at $x_0$.
Each simulation was initialized from a perturbed solid-liquid front in equilibrium and evolved for \(500~\mu\mathrm{s}\). Periodic boundary conditions were used in the transverse \(y\)-direction. In the \(x\)-direction, i.e., the growth direction, a no-flux condition was imposed on the solid-side boundary, while fixed liquid values (\(\phi=-1,\:c=c_\infty\)) were imposed on the opposite boundary to supply the incoming liquid in the moving-frame calculation. All simulations used the same grid spacing, \(\Delta x=0.8W\), and explicit time step, \(\Delta t/\tau_0=0.15(\Delta x/W)^2\). A grid-resolution check supporting this choice is reported in Supplementary Section~S5. The isotropic finite difference operations described in Ref.~\cite{Ji2022} were used for the Laplacian and divergence operations, and simple finite differences were used for the anisotropy terms~\cite{Tourret2015}.

\begin{figure*}[htp!]
\centering
\includegraphics[]{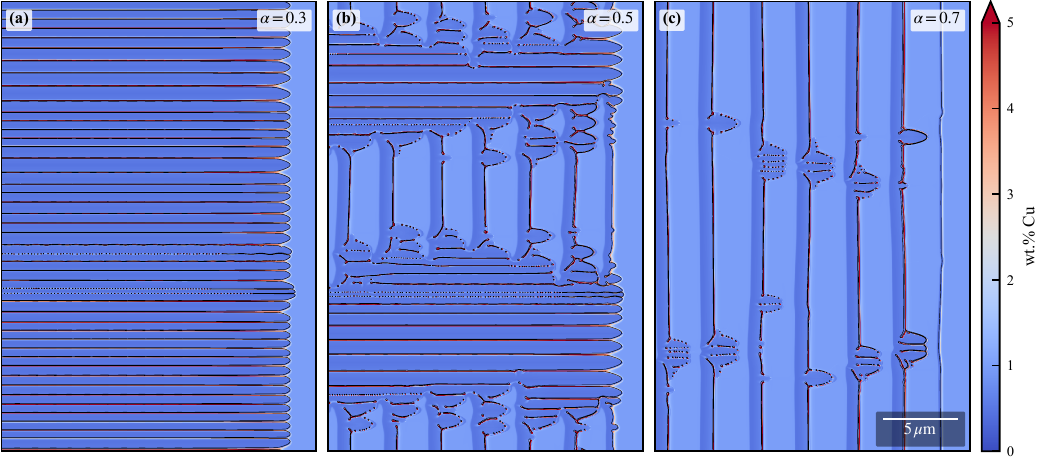}
\caption{
    Effect of solute drag parameter  in two-dimensional Al--Cu solidification simulations at pulling speed \(V_\mathrm{pull}=0.12~\mathrm{m\,s^{-1}}\) and thermal gradient \(G=5\times10^6~\mathrm{K\,m^{-1}}\). The \(\alpha=0.3\) case in (a) develops a cellular morphology, the \(\alpha=0.5\) case in (b) shows coexistence of banded and cellular features, and the \(\alpha=0.7\) case in (c) forms a predominantly banded structure. The solid black line denotes the \(\phi=0\) contour.
    }
    \label{fig:alcu_microstructures}
\end{figure*}

Figure~\ref{fig:alcu_microstructures} summarizes the resulting solidification morphologies for the three optimized drag responses with the imposed pulling velocity set to \(V_\mathrm{pull}=0.12~\mathrm{m\,s^{-1}}\). The lowest drag case, \(\alpha=0.3\), develops a cellular morphology over the simulated time window. Increasing the drag coefficient to \(\alpha=0.5\) produces a mixed response in which banded regions and cellular features coexist. This intermediate case is particularly informative because it suggests that the simulation is close to a transition between two competing growth modes. Finally, the high-drag case, \(\alpha=0.7\), evolves into a predominantly banded structure over the same time interval. The trend is therefore systematic: increasing the prescribed drag contribution promotes a transition from cellular growth toward banded growth under otherwise identical conditions. Additional simulations revealed that \(\alpha=0.4\) results in a cellular array similar to that in Figure~\ref{fig:alcu_microstructures}(a), while simulations with \(\alpha=0.6\) and \(0.8\) exhibit banded microstructures similar to Figure~\ref{fig:alcu_microstructures}(c).
As an additional check that the banded morphologies are not an artifact of the enlarged diffuse-interface width, we repeated a representative calculation using \(\alpha=0.7\) with a smaller interface upscaling factor, \(S=3\) with a re-optimized interpolation function and a smaller domain. The resulting simulation produced a qualitatively similar banded structure for the high-drag case and comparable band spacing to the \(S=5\) calculations. Thus, the observed banding is not simply a numerical artifact of the chosen interface width. Fine-scale cellular features should be interpreted more cautiously: the wavy intercellular channels visible in Figs.~\ref{fig:alcu_microstructures}(a) and (b) resemble features discussed by Mancias \textit{et al.}~\cite{Mancias2025}, who found that such waviness was removed when decreasing the interface upscaling factor to \(S=3\). This suggests that the small-scale channel morphology may be more sensitive to interface stretching than the large-scale cellular-to-banded transition; excess interfacial transport along the diffuse interface is one possible contributor~\cite{Ji2025}.

The mixed morphology observed for \(\alpha=0.5\) is consistent with a system close to the transition between cellular and banded growth. Over the simulated time window, banded regions and cellular features coexist rather than one morphology completely replacing the other. Longer simulations, different initial conditions, or additional noise terms could shift the balance toward a more clearly selected cellular or banded state, but the coexistence itself is informative: it indicates that the intermediate drag response places the system near the boundary between the two growth modes.

\begin{figure*}[htp!]
\centering
\includegraphics[]{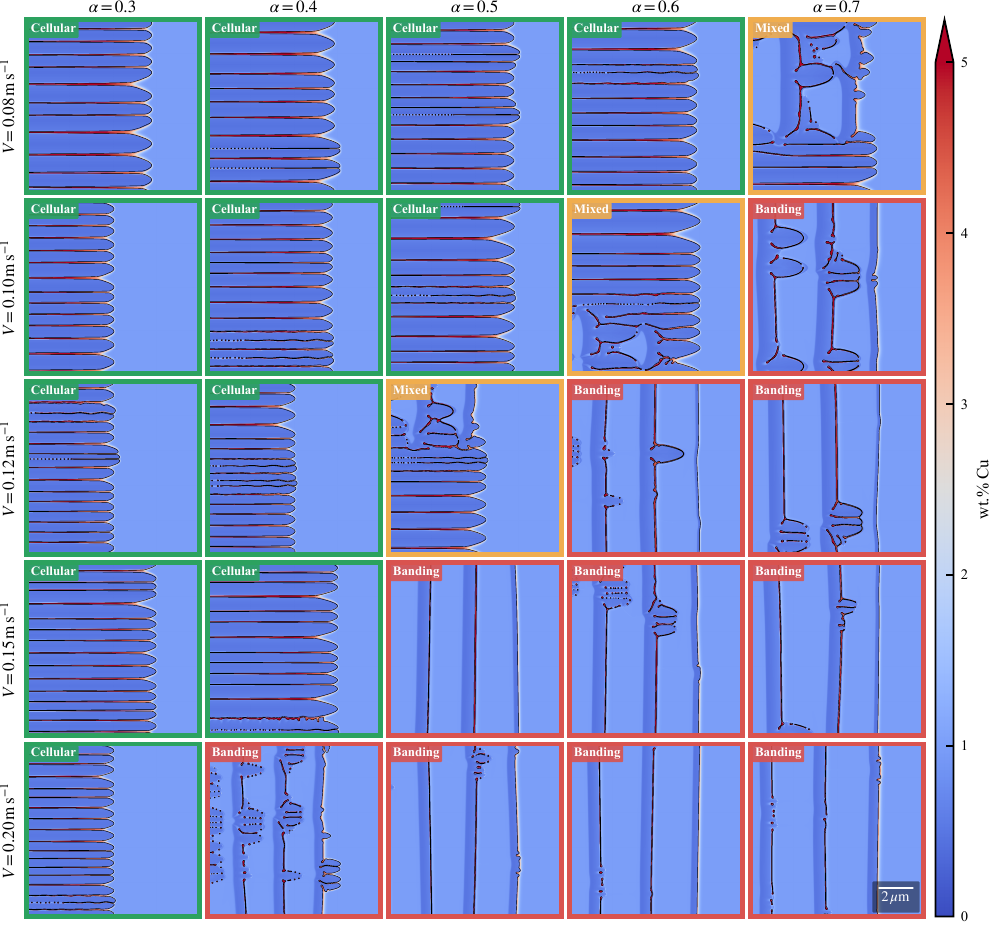}
\caption{
Concentration-field maps from the two-dimensional Al--Cu simulations over the investigated pulling-velocity and solute-drag parameter space. Rows correspond to \(V_{\mathrm{pull}}=0.08\), \(0.10\), \(0.12\), \(0.15\), and \(0.20~\mathrm{m\,s^{-1}}\), while columns correspond to \(\alpha=0.3\), \(0.4\), \(0.5\), \(0.6\), and \(0.7\). Each panel shows a \(10\times10~\mu\mathrm{m}^2\) crop of the concentration field. The simulations show a systematic progression from cellular through mixed to predominantly banded growth with increasing pulling velocity and solute-drag coefficient. The labels denote the dominant morphology observed within the simulated domain and time interval.}
    \label{fig:alcu_velocity_drag_matrix}
\end{figure*}

To determine whether this progression is specific to the selected pulling velocity, the simulations were extended to a matrix of five pulling velocities and five solute-drag coefficients, using the same simulation setup. Figure~\ref{fig:alcu_velocity_drag_matrix} summarizes the resulting concentration fields using \(10\times10~\mu\mathrm{m}^2\) crops from the corresponding simulations.

The data corresponding to the expanded matrix of parameters shown in Fig.~\ref{fig:alcu_velocity_drag_matrix} reveal that the progression observed in Fig.~\ref{fig:alcu_microstructures} forms part of a systematic trend over the investigated parameter range. At \(V_{\mathrm{pull}}=0.08~\mathrm{m\,s^{-1}}\), cellular growth persists up to \(\alpha=0.6\), with mixed features appearing at \(\alpha=0.7\). As the pulling velocity increases, mixed and banded growth appear at progressively smaller values of \(\alpha\). At \(V_{\mathrm{pull}}=0.20~\mathrm{m\,s^{-1}}\), banded structures are observed from \(\alpha=0.4\) onward. Thus, increasing either the pulling velocity or the solute-drag coefficient promotes banded growth within the investigated parameter range.

This systematic change in morphology suggests that the drag-modified kinetic response alters the stability of the laterally averaged front. We next interpret this trend using the kinetic solidus relation implied by the prescribed sharp-interface response functions.

\subsection{Kinetic-solidus interpretation of the drag-dependent solidification response}
\label{subsec:binary_alcu_2d}

To interpret the trend shown in Figures~\ref{fig:alcu_microstructures} and~\ref{fig:alcu_velocity_drag_matrix}, we first examine the kinetic liquidus and solidus branches implied by the target response functions. For a dilute binary alloy with nominal composition \(c_\infty\), the velocity-dependent branches used here are
\begin{align}
    T_L^\infty(V;\alpha)
    &=
    T_m
    -
    m(V;\alpha)c_\infty
    -
    \frac{V}{\mu_0},
    \label{eq:TL_inf_alcu}
    \\
    T_S^\infty(V;\alpha)
    &=
    T_m
    -
    \frac{m(V;\alpha)}{k(V)}c_\infty
    -
    \frac{V}{\mu_0}.
    \label{eq:TS_inf_alcu}
\end{align}
Here \(T_L^\infty\) and \(T_S^\infty\) do not describe the local interface temperature at the solid-liquid front; rather, they are branches of a kinetic phase diagram constructed from the prescribed rapid-solidification response functions. For the one-sided directional-solidification conditions used here, the solid composition incorporated behind the front is \(c_\infty\). The steady state mean front temperature therefore lies on the kinetic solidus branch \(T_S^\infty(V;\alpha)\), which is why the reduced growth-rate diagnostic below is based on the local slope \(dT_S^\infty/dV\). The separation of \(T_L^\infty\) and \(T_S^\infty\)  defines an effective kinetic freezing range of the alloy,
\begin{equation}
    \Delta T_{\rm fr}(V;\alpha)
    =
    T_L^\infty(V;\alpha)
    -
    T_S^\infty(V;\alpha)
    =
    m(V;\alpha)c_\infty
    \left[
        \frac{1}{k(V)}-1
    \right].
    \label{eq:kinetic_freezing_range_alcu}
\end{equation}
Since \(k(V)\) is common to all drag cases in the present comparison, the drag dependence of \(\Delta T_{\rm fr}\) enters primarily through \(m(V;\alpha)\).

\begin{figure}[t]
    \centering
    \includegraphics[]{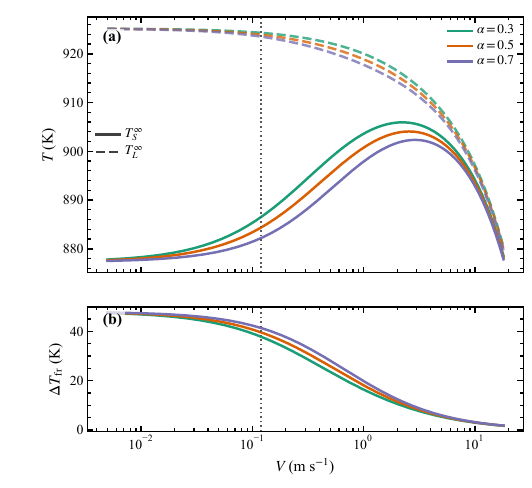}
    \caption{
    Drag-dependent kinetic phase-boundary context for the binary Al--Cu target response functions. (a) Kinetic solidus and liquidus branches, \(T_S^\infty(V;\alpha)\) and \(T_L^\infty(V;\alpha)\), constructed from the prescribed velocity-dependent partition coefficient and drag-modified kinetic liquidus response. Solid lines denote \(T_S^\infty\), dashed lines denote \(T_L^\infty\), and colors indicate the solute drag coefficient \(\alpha\). (b) Corresponding kinetic freezing interval, \(\Delta T_{\rm fr}=T_L^\infty-T_S^\infty\). The vertical dotted line marks the pulling velocity used in the two-dimensional simulations in Figure~\ref{fig:alcu_microstructures}}.
    \label{fig:alcu_kinetic_branches}
\end{figure}

Figure~\ref{fig:alcu_kinetic_branches} shows that increasing \(\alpha\) increases the kinetic freezing interval over the velocity range relevant to the simulations. This provides a first, direct indication of how the drag-modified liquidus response changes the rapid-solidification thermodynamics sampled by the front. However, the freezing interval alone does not explain why the morphology becomes more susceptible to instabilities with increasing drag. To examine this connection more directly, we perform a reduced stability analysis of the interface.
We write a small perturbation of the interface position in the standard normal-mode form
\begin{equation}
    X_f(y,t)
    =
    X_0
    +
    \delta X
    \exp\!\left[
        \sigma(Q)t+iQy
    \right],
    \label{eq:interface_perturbation_ansatz}
\end{equation}
where \(Q\) is the lateral perturbation wavenumber and \(\sigma(Q)\) is the corresponding growth rate. In a full morphological-stability calculation, \(\sigma(Q)\) would be obtained by solving the perturbed solute-transport problem together with the interfacial boundary conditions~\cite{MullinsSekerka1964,KGT1986,KarmaSarkissian1993}. Here, we isolate only the uniform mode \(Q=0\). This mode corresponds to a laterally uniform displacement of the mean planar front and therefore does not describe cellular or dendritic finite-wavelength perturbations. It nevertheless provides a compact measure of how the steady-state kinetic solidus relation converts a mean front displacement in the imposed thermal gradient into a perturbation of the mean interface velocity.

To proceed, we consider \(Q=0\), for which the perturbation is independent of \(y\), and the perturbed interface position is
\begin{equation}
    X_f(t)
    =
    X_0
    +
    \delta X \exp(\sigma_0 t),
    \label{eq:q0_interface_perturbation}
\end{equation}
where \(\sigma_0\equiv \sigma(Q=0)\). The corresponding perturbation of the interface velocity is obtained as
\begin{equation}
    \delta V
    =
    \frac{d}{dt}
    \left[
        \delta X \exp(\sigma_0 t)
    \right]
    =
    \sigma_0
    \delta X
    \exp(\sigma_0 t).
    \label{eq:q0_velocity_perturbation}
\end{equation}
Equivalently, since the instantaneous front displacement is
\(\delta X_f(t)=\delta X\exp(\sigma_0 t)\), Eq.~\eqref{eq:q0_velocity_perturbation} gives
\begin{equation}
    \delta V
    =
    \sigma_0 \delta X_f .
    \label{eq:q0_kinematic_relation}
\end{equation}

The imposed thermal gradient converts the same front displacement into a local temperature perturbation,
\begin{equation}
    \delta T
    =
    G\,\delta X_f ,
    \label{eq:deltaT_uniform_front}
\end{equation}
with the sign convention chosen so that a positive displacement samples a temperature increase \(G\delta X_f\). If the perturbed mean front remains close to the steady kinetic solidus relation, the corresponding temperature and velocity perturbations are related by the local linearization
\begin{equation}
    \delta T
    =
    \frac{dT_S^\infty}{dV}\,\delta V .
    \label{eq:branch_linearization}
\end{equation}
Combining Eqs.~\eqref{eq:q0_kinematic_relation}--\eqref{eq:branch_linearization} gives
\begin{equation}
    G\,\delta X_f
    =
    \frac{dT_S^\infty}{dV}
    \sigma_0 \delta X_f .
\end{equation}
For a nonzero perturbation amplitude, this yields
\begin{equation}
    \sigma_0
    =
    \frac{G}{dT_S^\infty/dV}.
    \label{eq:sigma0_reduced_growth_rate}
\end{equation}
Hence, \(\sigma_0\) is a reduced \(Q=0\) response defined from the local slope of the steady drag-modified kinetic-solidus relation \(T_S^\infty(V;\alpha)\). It measures how a laterally uniform front displacement in the imposed thermal gradient is converted, through the local slope \(dT_S^\infty/dV\), into a perturbation of the mean interface velocity. Because \(dT_S^\infty/dV>0\) for all cases considered here, \(\sigma_0\) is positive for every drag coefficient and is not used as a sign-based stability criterion. Its magnitude is the relevant indicator: larger \(\sigma_0\) corresponds to a more responsive mean front. The full two-dimensional phase field dynamics determine whether this laterally averaged tendency is preempted by finite-wavelength dendritic/cellular growth or develops into relaxation-type banding. The purpose of \(\sigma_0\) is therefore to isolate the laterally averaged response associated with the kinetic solidus branch, which is relevant to relaxation-type banding and oscillations~\cite{KarmaSarkissian1993}. 

The slope entering Eq.~\eqref{eq:sigma0_reduced_growth_rate} can be solved for explicitly from Eq.~\eqref{eq:TS_inf_alcu} as
\begin{equation}
    \frac{dT_S^\infty}{dV}
    =
    -
    c_l
    \frac{dm}{dV}
    +
    m
    \frac{c_\infty}{k^2}
    \frac{dk}{dV}
    -
    \frac{1}{\mu_0},
    \label{eq:dTSdV_decomposition}
\end{equation}
where \(c_l=c_\infty/k(V)\), \(k=k(V)\), and \(m=m(V;\alpha)\). This decomposition shows how the drag-modified liquidus response enters the reduced \(Q=0\) growth-rate estimate. The term proportional to \(dk/dV\) is associated with the velocity dependence of solute trapping, whereas the term proportional to \(dm/dV\) captures the velocity sensitivity of the drag-modified kinetic liquidus response. Although \(k(V)\) and \(dk/dV\) are independent of \(\alpha\), both solutal terms depend on \(\alpha\) through \(m(V;\alpha)\). Evaluation of the complete derivative shows that their net effect is to reduce the positive slope \(dT_S^\infty/dV\) with increasing \(\alpha\) over the velocity range considered here. Consequently, \(\sigma_0=G/(dT_S^\infty/dV)\) increases with \(\alpha\).
\begin{figure}[ht!]
    \centering
    \includegraphics[]{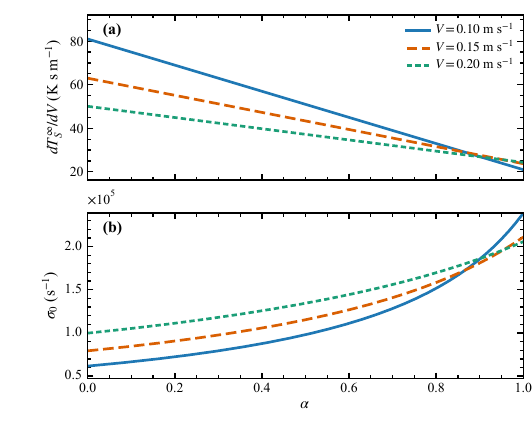}
    \caption{
    Reduced \(Q=0\) growth-rate diagnostics evaluated at three pulling velocities used in the two-dimensional simulations. (a) Slope of the kinetic solidus relation \(dT_S^\infty/dV\) as a function of the solute drag coefficient. (b) Corresponding reduced \(Q=0\) growth-rate estimate \(\sigma_0=G/(dT_S^\infty/dV)\).
    }
    \label{fig:alcu_branch_feedback}
\end{figure}
Figure~\ref{fig:alcu_branch_feedback} shows this trend quantitatively when evaluated at three representative velocities used in the simulations. Over the range \(0\le\alpha\le1\), increasing \(\alpha\) decreases the slope \(dT_S^\infty/dV\) and increases the reduced \(Q=0\) growth-rate estimate \(\sigma_0\). In physical terms, a given uniform displacement of the interface in the imposed thermal gradient corresponds to a larger perturbation of the mean front velocity when the drag contribution is larger. This reduced \(Q=0\) result therefore indicates that increasing \(\alpha\) strengthens the laterally averaged response associated with temporal front oscillations. It does not, however, determine whether this mode dominates over finite-wavelength cellular or dendritic perturbations, which would require a complete stability analysis~\cite{KarmaSarkissian1993}. The increasing \(\sigma_0\) should consequently be interpreted as a diagnostic consistent with the greater prevalence of relaxation-type banding in the high-drag simulations, instead of as an independent prediction of the selected morphology.

\subsection{Dilute ternary Al--Si--Cu alloy}

As a representative demonstration of the dilute multicomponent response-matching formulation, we consider directional solidification of a dilute Al--1 wt.\% Si--1 wt.\% Cu alloy. The purpose of this example is not to construct a complete Al--Si--Cu microstructure map, but to demonstrate that independently specified solute response functions determined from the optimization methodology derived in this work can be embedded simultaneously in a two-dimensional phase field simulation. The Al--Cu parameters are the same as in the binary calculations, except that the nominal Cu concentration is set to \(c_{\infty,\mathrm{Cu}}=1~\mathrm{wt.\%}\). For Si, we use \(k_{\mathrm{Si}}=0.13\), \(m_{\mathrm{Si}}=6.5~\mathrm{K\,wt.\%^{-1}}\), and \(D_{l,\mathrm{Si}}=5.5\times 10^{-9}~\mathrm{m^2\,s^{-1}}\) from Ref.~\cite{Zhong2025_nature}. Compared to the Al--Cu system, the ratio \(m_i/(1-k_i)\) differs by a factor of \(\sim2.5\) for the two dilute species considered here. The corresponding diffusive speed is estimated as \(V_{D,\mathrm{Si}}\sim D_{l,\mathrm{Si}}/W_0=5.5~\mathrm{m\,s^{-1}}\). The Gibbs--Thomson coefficient is taken as \(\Gamma=2.18\times10^{-7}~\mathrm{K\,m}\), obtained by averaging the values used in Refs.~\cite{Ji2023,Zhong2025_nature} for binary Al--Cu and Al--Si systems. A common drag coefficient \(\alpha=0.4\) is used for both solutes, with individually optimized diffusion interpolation functions.

\begin{figure}
    \centering
    \includegraphics[]{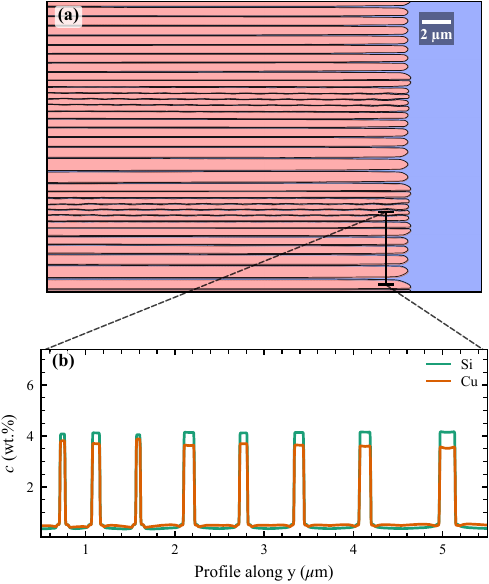}
    \caption{ Dilute ternary Al--Si--Cu directional-solidification demonstration. (a) Order-parameter field from the Al--1 wt.\% Si--1 wt.\% Cu calculation, showing the cellular morphology obtained with simultaneous Si and Cu redistribution. The solid black line denotes the \(\phi=0\) contour and the highlighted line marks a representative section crossing several liquid channels. (b) Si and Cu concentration profiles along the highlighted section in (a), used to compare the solute-specific segregation response within the same evolving morphology.}
    \label{fig:ternary_al_sicu}
\end{figure}

Figure~\ref{fig:ternary_al_sicu} summarizes a ternary moving-frame calculation with \(V_\mathrm{pull}=0.10~\mathrm{m\,s^{-1}}\) and \(G=5\times10^6~\mathrm{K\,m^{-1}}\) and a \(10\times 20~\mu\mathrm{m}^2\) domain.
Rather than showing the two concentration fields separately, we use the order-parameter field to show the full cellular morphology and compare the Si and Cu redistribution along a representative line section crossing several cells and liquid channels. This presentation emphasizes the main purpose of the calculation: both solutes are transported by the same evolving solid-liquid interface, while their segregation amplitudes and local profile shapes remain controlled by their independently specified thermodynamic and transport parameters. The result therefore provides a direct two-dimensional demonstration of the dilute multicomponent model, in which multiple solute response functions are emulated in the same enlarged-interface phase field simulation without reducing the problem to a quasi-binary alloy system~\cite{Pinomaa2020_316L,Kroportin2023_inconel718,Mancias2025}.
The line profiles in Fig.~\ref{fig:ternary_al_sicu}(b) make the comparison between the two solutes clearer than the corresponding two-dimensional concentration maps, which are visually similar because they follow the same cellular liquid-channel network. The Si and Cu profiles are qualitatively similar, as expected because the two dilute solutes have comparable properties and are redistributed by the same evolving interface morphology. Their concentration levels nevertheless differ both in the solid and liquid regions, illustrating the multicomponent capability of the formulation: the model evolves multiple solute fields with independently specified partitioning, liquidus response, and transport parameters, and therefore predicts species-specific segregation within the same cellular microstructure. The profiles also show weak shoulders near some liquid-channel concentration maxima. Since these shoulders occur close to the diffuse solid-liquid interfaces, they may be affected by the enhanced interfacial diffusivity used to recover the target trapping response and are expected to occur with any number of components. Thus, the ternary calculation demonstrates simultaneous multicomponent response-function embedding, while also highlighting that quantitative channel-scale microsegregation metrics may require additional convergence checks regarding enhanced interfacial transport~\cite{Ji2025}.

\section{Discussion} \label{sec:discussion} The central contribution of this work is a response-function mapping strategy that embeds prescribed sharp-interface rapid-solidification kinetics into an enlarged-interface phase field model. The binary and multicomponent calculations are therefore used primarily to demonstrate the capabilities and consequences of the proposed methodology, rather than to provide a comprehensive investigation of microstructure selection. In particular, the two-dimensional calculations illustrate why quantitative control of the velocity-dependent interface response can be important even when the underlying thermodynamic and transport parameters are otherwise unchanged.

Traditionally, phase field calculations with enlarged-interfaces have employed interfacial diffusivity interpolations that are usually chosen from a restricted analytical family, often with one or a few adjustable parameters. Such choices can reproduce selected asymptotic limits or qualitative kinetic trends, and they have been essential for making rapid-solidification phase field simulations computationally feasible. However, they do not generally provide robust quantitative control over the full velocity dependence of both solute trapping and the kinetic liquidus response. 

Here, the interpolation function is instead treated as a constrained response-matching degree of freedom. The diffuse-interface model is calibrated against prescribed sharp-interface functions \(k_i(V)\) and \(m_i(V;\alpha)\), and the accepted interpolation is required to satisfy explicit accuracy tolerances over the velocity interval of interest. This distinction is important for rapid solidification, where microstructure selection depends not only on equilibrium alloy parameters, but also on the velocity dependence of partitioning, kinetic undercooling, solute drag, and the resulting phase-boundary response~\cite{Tourret2023}. The objective of the calibration is consequently not to select a convenient analytical form for the interfacial diffusivity, but to ensure that the enlarged-interface model quantitatively reproduces prescribed sharp-interface kinetics over a relevant velocity interval.

The binary Al--Cu calculations demonstrate our approach in a deliberately controlled setting. The target partition coefficient was kept identical for all drag coefficients, while only the drag-dependent kinetic liquidus response was varied. The extended investigation over a $5 \times 5$ matrix of velocity and drag values further shows that over the investigated range, increasing either \(V_{\mathrm{pull}}\) or \(\alpha\) promotes the appearance of banded growth, and the value of \(\alpha\) at which banding is observed decreases as the pulling velocity increases. Because the nominal composition, thermal gradient, and prescribed solute-trapping response remain common across the simulations, these results demonstrate that quantitative changes in the drag-modified kinetic liquidus response can materially influence the predicted morphology. This is the primary role of the microstructure calculations in the present methods study: they show that accurate embedding of the kinetic response is not merely a one-dimensional calibration exercise, but can affect the outcomes of multidimensional simulations. 

The reduced \(Q=0\) calculation provides a complementary interpretation of the mean-front response over the same velocity range. Increasing \(\alpha\) reduces the positive slope \(dT_S^\infty/dV\) and increases the reduced estimate \(\sigma_0=G/(dT_S^\infty/dV)\) at the pulling velocities considered in the two-dimensional calculations. This indicates that a given uniform displacement of the interface in the imposed thermal gradient corresponds to a larger perturbation of the laterally averaged front velocity when the drag contribution is larger. The calculation is not a finite-wavelength stability theory and does not predict the competition between cellular and banded modes, the band spacing, or dendrite-tip dynamics. Instead, it isolates the mean-front sensitivity associated with the kinetic solidus relation. Its consistency with the observed increase in banding tendency supports the interpretation that the drag-modified response contributes to the prevalence of relaxation-type oscillations, while the actual morphology remains determined by the full two-dimensional phase field dynamics.

The scalar drag parameter used here should, however, be interpreted with some care. In the continuous-growth-model description, \(\alpha\) controls the fraction of the solute-redistribution contribution that is subtracted from the driving force available for crystallization. This parameterization is useful for constructing tunable response functions, but it is not a complete microscopic theory of solute drag~\cite{Humadi2016}. Classical treatments, sharp-interface theories, diffuse-interface formulations, and atomistic studies differ in how the interfacial composition, dissipated free energy, and effective drag contribution should be defined~\cite{Cahn1962,HillertSundman1977_drag,Hillert1999_drag,AzizBoettinger1994_drag,Hareland2022}. Molecular dynamics and microscopic studies support the relevance of partial drag, but also indicate that its magnitude is alloy and velocity dependent rather than universal~\cite{Yang2011_drag_MD,Humadi2016,Kavousi2020_drag,Antillon2023,Haapalehto2022,Cui2026_drag_scripta,Cui2026_drag_acta_MD}. In the present framework, the drag response is therefore treated as an input response function to be tested or calibrated, not as a quantity fixed uniquely by the diffuse-interface interpolation. The range of \(\alpha\) considered in the microstructure calculations should accordingly be understood as a controlled sensitivity study of the prescribed kinetic response.

A related limitation is the choice of the continuous growth model as the target response. The CGM provides a compact and widely used benchmark for velocity-dependent partitioning and drag-modified liquidus kinetics, but it should not be regarded as a complete physical description at arbitrarily high velocities. Moreover, an alternative model presented recently by Hareland \textit{et al.} predicts that solute drag affects both the temperature and composition responses, resulting in a drag-dependent partition coefficient \(k(V;\alpha)\)~\cite{Hareland2022}. Such coupled response functions can be incorporated directly into the present framework by using \(k(V;\alpha)\) and \(m(V;\alpha)\) as the simultaneous optimization targets. In the original CGM, complete solute trapping is approached asymptotically as \(V\) increases, whereas local-nonequilibrium models predict complete trapping at a finite velocity by relaxing the assumption that the diffusive flux adjusts instantaneously~\cite{Sobolev1995,Sobolev1997,Galenko2007}. Atomistic simulations and recent nonequilibrium interface models similarly motivate response functions that differ from the asymptotic CGM form at high velocity~\cite{Yang2011_drag_MD,Haapalehto2022}. This limitation concerns the selected demonstration targets rather than the calibration strategy itself. The proposed method is response-function agnostic: CGM targets are used here as a transparent demonstration case, but response functions inferred from any local nonequilibrium models, atomistic simulations or rapid solidification experiments could be embedded in the same way, provided that they can be represented by smooth and numerically stable interpolation functions. 

The dilute multicomponent extension of the phase field model should be viewed in the same spirit. Within the diagonal independent-solute approximation, each solute can be assigned its own equilibrium partition coefficient, liquidus slope, diffusive speed, and target trapping response. The use of Eq.~(\ref{eq:dilute_helmholtz_density}) to describe the free energy of each phase removes the thermodynamic restriction that appears when a universal ideal-dilute prefactor is used for all solutes, which provides a controlled route to multicomponent response-function matching. The ternary calculation consequently serves as a demonstration that several prescribed solute responses can be embedded simultaneously and transferred to a multidimensional simulation.

However, concentrated or strongly interacting alloys generally require thermodynamic and kinetic coupling beyond independent scalar diffusivities. In such systems, redistribution occurs through a diffusion matrix, and cross diffusion can affect both the rate and direction of interfacial redistribution in composition space~\cite{WangTourret2024_diffusion}. Since solute drag is associated with dissipation during redistribution, cross-coupled transport may also modify the effective drag response in ways that cannot be captured by independent binary-like parameters. Recent multicomponent sharp-interface treatments emphasize the importance of non-ideal phase diagrams, velocity-dependent phase-boundary properties, and full liquid diffusion matrices~\cite{Hareland2022,Martin2024_drag}. Coupling the present response-function calibration to CALPHAD-based nonequilibrium phase diagrams and multicomponent diffusion data is therefore a natural next step. 

Several numerical limitations remain in the response function optimization methodology presented in this work. The additive Chebyshev representation with fixed thermodynamic interpolation performs well for the intermediate drag coefficients studied here, but becomes less robust for very small or very large \(\alpha\). In those regimes, matching both \(k(V)\) and \(m(V;\alpha)\) may require sharper features in \(q(\phi)\), or additional degrees of freedom beyond the diffusivity interpolation alone. The tolerance study performed illustrates this tradeoff: overly strict matching can introduce unnecessary structure into \(q(\phi)\), whereas overly loose matching reduces quantitative control over the response functions. The \(1\%\) tolerance used in the production calculations was selected as a practical compromise between response accuracy and smoothness. Future implementations could optimize the thermodynamic interpolation \(g(\phi)\) together with \(q(\phi)\). Alternative constrained representations could also be employed when the target responses cannot be reproduced robustly by the present parameterization.

Finally, the two-dimensional {microstructure} simulations should be interpreted as controlled demonstrations of the consequences of response-function calibration rather than as a complete processing or stability map. The expanded matrix establishes a systematic dependence of the observed morphology on \(V_{\mathrm{pull}}\) and \(\alpha\), but the classifications refer to finite simulation domains and time intervals and do not define sharp asymptotic stability boundaries. A {future} quantitative study of banding thresholds should employ experimentally or atomistically derived response functions, a more finely resolved parameter space, and interpolation functions calibrated over the complete velocity range sampled by the evolving interfaces~\cite{SmithAziz1994_Al,Haapalehto2022}.

Enhanced interfacial diffusivity can introduce excess surface diffusion along the diffuse interface, which is distinct from the normal redistribution kinetics targeted by the optimization. Previous rapid solidification phase field studies found this effect to be small in some dendrite-growth tests, but its interaction with interface stretching, high-velocity dynamics, and finite-wavelength morphology selection remains an important numerical issue~\cite{Ji2025}. In the present calculations, reducing the interface upscaling factor produced qualitatively similar banded structures and comparable band spacing, supporting the conclusion that the observed banding trend is not an artifact of the enlarged interface width. 

Nevertheless, quantitative predictions of channel-scale microsegregation, band spacing, dendrite-tip dynamics, and stability boundaries will require additional interface-width and resolution studies, longer simulations, and eventually three-dimensional calculations~\cite{nanotoday} with thermal diffusion~\cite{Song2018_TFC}. The microstructure results therefore support, rather than redefine, the methodological contribution of this work. The demonstrated sensitivity to solute drag highlights why nonequilibrium interface kinetics should be represented quantitatively and transparently rather than being determined implicitly by a convenient diffusivity interpolation.

\section{Conclusions} \label{sec:conclusions} 

We developed an optimization-based phase field framework for alloys that makes it possible to prescribe velocity-dependent rapid-solidification response functions in enlarged-interface simulations. The interfacial diffusivity interpolation was treated as a constrained calibration function and optimized so that one-dimensional steady phase field solutions reproduce target responses for both the velocity-dependent partition coefficient and the drag-modified kinetic liquidus response. The central methodological advance is the replacement of an assumed analytical interpolation by an explicitly validated response-matching procedure.

Applied to dilute Al--Cu, the optimized additive Chebyshev representation of the interface diffusivity function reproduced the prescribed continuous-growth-model responses for intermediate solute drag coefficients. The tolerance study showed that a \(1\%\) target provides a useful compromise between response-function accuracy and smoothness of the optimized interpolation through the interface. The optimized diffusivity profile was also effectively converged with the \(N=8\) representation used in the production calculations.

The consequences of the calibrated response functions were demonstrated in two-dimensional directional-solidification simulations spanning five pulling velocities and five drag coefficients. Increasing either \(V_{\mathrm{pull}}\) or \(\alpha\) promoted a progression from cellular through mixed to predominantly banded growth over the investigated range, with banding appearing at lower \(\alpha\) as the pulling velocity increased. A reduced \(Q=0\) analysis over the same velocity range showed a corresponding increase in the sensitivity of the laterally averaged front with increasing \(\alpha\). This diagnostic is consistent with the observed banding tendency but does not independently predict the selected morphology. These simulations serve primarily to demonstrate that quantitative differences in the prescribed kinetic response can materially affect multidimensional phase field predictions, rather than to provide a comprehensive investigation of microstructure selection.

Finally, we also extended the formulation to dilute multicomponent alloys in a form that permits independent specification of equilibrium partition coefficients and liquidus slopes for each solute species. The ternary calculation demonstrated that several prescribed solute responses can be embedded simultaneously and transferred to multidimensional simulations without additional fitting. The framework therefore provides a practical route for embedding prescribed nonequilibrium interface kinetics into quantitative phase field simulations of rapid solidification in binary and multicomponent alloys. 

More broadly, the framework makes uncertain nonequilibrium interface physics explicit in mesoscale simulations. The present model can be calibrated to emulate response functions obtained from sharp-interface theory, atomistic simulations, or experiments over a prescribed velocity range. Further development should focus on coupled multicomponent thermodynamics and transport, calibration to non-CGM kinetic responses, and the numerical convergence required for quantitative predictions of microsegregation and morphology selection.

\section*{CRediT authorship contribution statement}

\textbf{Joni Kaipainen:} Writing - Original Draft, Software, Conceptualization, Methodology, Investigation, Visualization, Formal analysis.

\textbf{Tatu Pinomaa:} Writing - Review \& Editing, Conceptualization, Supervision, Project administration, Funding acquisition.

\textbf{Nikolas Provatas:} Writing - Review \& Editing, Conceptualization, Formal analysis, Supervision, Project administration.

\section*{Declaration of competing interest}

The authors declare that they have no known competing financial interests or personal relationships that could have appeared to influence the work reported in this paper.

\section*{Acknowledgments}

The work of J.K. and T.P. was supported by the Research Council of Finland through Grant No. 362197. N.P. thanks The Natural Sciences and Engineering Research Council (NSERC) of Canada and Canada Research Chairs (CRC) Program for funding. The authors acknowledge CSC—IT Center for Science, Finland, for computational resources.

\bibliographystyle{elsarticle-num} 
\bibliography{references}

\clearpage
\onecolumn

\section*{Supplementary Material}
\setcounter{section}{0}
\setcounter{equation}{0}
\setcounter{figure}{0}
\setcounter{table}{0}

\renewcommand{\thesection}{S\arabic{section}}
\renewcommand{\thesubsection}{S\arabic{section}.\arabic{subsection}}
\renewcommand{\theequation}{\thesection.\arabic{equation}}
\renewcommand{\thefigure}{S\arabic{figure}}
\renewcommand{\thetable}{S\arabic{table}}

\numberwithin{equation}{section}
\renewcommand{\theequation}
    {S\arabic{section}.\arabic{equation}}

\section{Restriction imposed by the ideal dilute prefactor}
\label{app:ideal_prefactor_restriction}

This Supplementary Section shows why a direct multicomponent extension of the ideal dilute Helmholtz free-energy density used in binary alloy phase field models cannot, in general, prescribe independent pairs of equilibrium partition coefficients and liquidus slopes for multiple solutes. The restriction is not caused by the use of a Helmholtz free energy itself. It arises when all solute species are assigned the same ideal dilute thermodynamic prefactor. In that case, the partition coefficients fix the solute free-energy jumps, and the same prefactor then determines the liquidus slopes. The result is a constraint between \(k_i\) and \(m_i\) that does not appear in binary alloys but becomes restrictive in ternary and higher-order dilute alloys.

\subsection{Equilibrium partitioning}
\label{app:ideal_partitioning}

Consider one solvent and \(n_s\) dilute solute species indexed by \(i=1,\ldots,n_s\). The direct ideal dilute Helmholtz form for a bulk phase \(\alpha=s,l\) is
\begin{equation}
    f^\alpha
    =
    f_0^\alpha(T)
    +
    \sum_{i=1}^{n_s}
    \left[
        A_0(c_i^\alpha \ln c_i^\alpha-c_i^\alpha)
        +
        \epsilon_i^\alpha c_i^\alpha
    \right],
    \label{eq:app_ideal_helmholtz_bulk}
\end{equation}
where \(f_0^\alpha(T)\) is the pure-solvent contribution, \(\epsilon_i^\alpha\) is the dilute solute free-energy offset in phase \(\alpha\), and
\begin{equation}
    A_0=\frac{RT_m}{v_0}
\end{equation}
is taken to be the same for all solutes.

The diffusion potential of solute \(i\) is
\begin{equation}
    \mu_i^\alpha
    =
    \frac{\partial f^\alpha}{\partial c_i^\alpha}
    =
    A_0 \ln c_i^\alpha+\epsilon_i^\alpha.
    \label{eq:app_ideal_mu}
\end{equation}
At equilibrium across a flat solid-liquid interface,
\begin{equation}
    \mu_i^s=\mu_i^l,
\end{equation}
so that
\begin{equation}
    A_0\ln c_i^s+\epsilon_i^s
    =
    A_0\ln c_i^l+\epsilon_i^l .
\end{equation}
Therefore
\begin{equation}
    A_0\ln\left(\frac{c_i^s}{c_i^l}\right)
    =
    -(\epsilon_i^s-\epsilon_i^l).
\end{equation}
Defining the equilibrium partition coefficient
\begin{equation}
    k_i=\frac{c_i^s}{c_i^l},
\end{equation}
and the solute free-energy jump
\begin{equation}
    \Delta\epsilon_i=\epsilon_i^s-\epsilon_i^l,
\end{equation}
gives
\begin{equation}
    \Delta\epsilon_i
    =
    -A_0\ln k_i.
    \label{eq:app_ideal_delta_epsilon}
\end{equation}
Thus the ideal dilute Helmholtz form can reproduce arbitrary prescribed partition coefficients \(k_i\) by choosing the corresponding solute free-energy jumps. The restriction appears only when the liquidus hypersurface is considered.

\subsection{Liquidus hypersurface}
\label{app:ideal_liquidus}

The phase grand-potential density corresponding to Eq.~\eqref{eq:app_ideal_helmholtz_bulk} is
\begin{equation}
    \omega^\alpha
    =
    f^\alpha-\sum_i \mu_i c_i^\alpha .
\end{equation}
Using Eq.~\eqref{eq:app_ideal_mu}, this becomes
\begin{align}
    \omega^\alpha
    &=
    f_0^\alpha(T)
    +
    \sum_i
    \left[
        A_0(c_i^\alpha\ln c_i^\alpha-c_i^\alpha)
        +
        \epsilon_i^\alpha c_i^\alpha
    \right] \nonumber\\
    &\qquad-
    \sum_i
    \left[
        A_0\ln c_i^\alpha+\epsilon_i^\alpha
    \right]c_i^\alpha
    \nonumber\\
    &=
    f_0^\alpha(T)
    -
    A_0\sum_i c_i^\alpha .
    \label{eq:app_ideal_grand_potential}
\end{align}
Bulk solid-liquid coexistence requires
\begin{equation}
    \omega^s=\omega^l .
\end{equation}
Using Eq.~\eqref{eq:app_ideal_grand_potential}, this gives
\begin{equation}
    f_0^s(T)-f_0^l(T)
    =
    A_0\sum_i(c_i^s-c_i^l).
    \label{eq:app_ideal_coexistence}
\end{equation}
Near the melting point of the pure solvent, we linearize
\begin{equation}
    f_0^s(T)-f_0^l(T)
    =
    \Delta f_T(T-T_m),
    \label{eq:app_ideal_delta_fT}
\end{equation}
where \(\Delta f_T\) is the pure-solvent thermal free-energy coefficient. Since \(c_i^s=k_i c_i^l\), Eq.~\eqref{eq:app_ideal_coexistence} becomes
\begin{equation}
    \Delta f_T(T-T_m)
    =
    -A_0\sum_i(1-k_i)c_i^l .
\end{equation}
Therefore the liquidus hypersurface predicted by the direct ideal dilute extension is
\begin{equation}
    T-T_m
    =
    -
    \sum_i
    \frac{A_0(1-k_i)}{\Delta f_T}
    c_i^l .
    \label{eq:app_ideal_liquidus}
\end{equation}
The desired local dilute liquidus hypersurface is
\begin{equation}
    T-T_m
    =
    -
    \sum_i m_i c_i^l ,
    \label{eq:app_target_liquidus}
\end{equation}
where \(m_i\) is the positive magnitude of the equilibrium liquidus slope of solute \(i\). Comparing Eqs.~\eqref{eq:app_ideal_liquidus} and \eqref{eq:app_target_liquidus} shows that the direct ideal dilute extension predicts
\begin{equation}
    m_i
    =
    \frac{A_0(1-k_i)}{\Delta f_T}.
    \label{eq:app_ideal_mi}
\end{equation}
Thus
\begin{equation}
    \frac{m_i}{1-k_i}
    =
    \frac{A_0}{\Delta f_T}
\end{equation}
is constrained to have the same value for every solute species. This is the restriction imposed by the universal ideal dilute prefactor. The model can prescribe arbitrary \(k_i\), but after the \(k_i\) have been fixed, the relative liquidus slopes are no longer independent. In a binary alloy this constraint is not visible because there is only one solute. In a multicomponent alloy, it restricts the set of dilute thermodynamic data that can be represented.

\subsection{Appearance of the restriction through the Gibbs-Thomson solvability condition}
\label{app:ideal_GT_restriction}

The same restriction appears when the one-dimensional phase field solvability condition is matched to the sharp-interface Gibbs-Thomson relation. This provides a useful consistency check and shows that the restriction is not an artifact of the bulk coexistence calculation.

For the direct ideal dilute Helmholtz extension, the solute offset across the diffuse interface may be written as
\begin{equation}
    \epsilon_i(\phi)
    =
    \bar{\epsilon}_i
    +
    \frac{\Delta\epsilon_i}{2}g(\phi),
\end{equation}
where \(g(-1)=-1\), \(g(+1)=+1\), and \(\Delta\epsilon_i=-A_0\ln k_i\). Defining
\begin{equation}
    b_i=\frac{1}{2}\ln k_i,
\end{equation}
one obtains
\begin{equation}
    \epsilon_i'(\phi)
    =
    -A_0 b_i g'(\phi).
\end{equation}
The corresponding dimensionless thermodynamic coupling is
\begin{equation}
    \lambda_i^{\rm id}
    =
    -\frac{A_0 b_i}{h},
    \label{eq:app_lambda_ideal}
\end{equation}
where \(h\) is the double-well barrier height.

For a one-dimensional interface moving with velocity \(V\), the integrated order-parameter equation has the generic form
\begin{equation}
    T
    =
    T_m
    -
    \sum_i
    \frac{\lambda_i^{\rm id}}{\Lambda_{\rm id}}
    I_i(V)
    +
    \frac{a_1^0W}{\Lambda_{\rm id}}\kappa
    +
    \frac{\tau_0a_1^0}{\Lambda_{\rm id}W}V,
    \label{eq:app_ideal_temperature_response}
\end{equation}
where
\begin{equation}
    I_i(V)
    =
    \int_{-\infty}^{+\infty}
    g'(\phi)c_i
    \frac{d\phi}{dx}\,dx,
\end{equation}
and
\begin{equation}
    \Lambda_{\rm id}
    =
    \frac{1}{M_0}
    \sum_i
    \lambda_i^{\rm id} I_i^0 .
\end{equation}
The equilibrium value of the integral is
\begin{equation}
    I_i^0
    =
    \frac{c_i^{l,0}(1-k_i)}{b_i}.
\end{equation}
In the equilibrium limit, matching the compositional part of Eq.~\eqref{eq:app_ideal_temperature_response} to the target Gibbs-Thomson relation requires
\begin{equation}
    m_i c_i^{l,0}
    =
    \frac{\lambda_i^{\rm id}}{\Lambda_{\rm id}}
    I_i^0 .
\end{equation}
Substituting Eqs.~\eqref{eq:app_lambda_ideal} and the expression for \(I_i^0\) gives
\begin{equation}
    m_i
    =
    -\frac{A_0}{h\Lambda_{\rm id}}(1-k_i).
\end{equation}
Hence
\begin{equation}
    \frac{m_i}{1-k_i}
    =
    -\frac{A_0}{h\Lambda_{\rm id}},
\end{equation}
which is common to all solutes. The one-dimensional solvability condition therefore recovers the same restriction obtained from the equilibrium liquidus hypersurface.

\section{Diagonal-susceptibility dilute thermodynamic construction}
\label{app:diagonal_susceptibility}

This Supplementary Section derives the thermodynamic construction used in the main text. The purpose is to replace the universal ideal dilute prefactor by component-specific diagonal susceptibilities while retaining a dilute independent-solute form. The derivation is written using a grand-potential representation, but the final equations can be expressed in concentration variables because the diagonal dilute relation between \(c_i\) and \(\mu_i\) is algebraically invertible.

\subsection{General grand-potential identities and diagonal approximation}
\label{app:general_gp_identities}

In a general multicomponent grand-potential description, each bulk phase \(\alpha=s,l\) is described by
\begin{equation}
    \omega^\alpha
    =
    \omega^\alpha(\boldsymbol{\mu},T),
\end{equation}
where \(\boldsymbol{\mu}=(\mu_1,\ldots,\mu_{n_s})\) are diffusion potentials. The phase concentrations are obtained as
\begin{equation}
    c_i^\alpha
    =
    -\frac{\partial \omega^\alpha}{\partial \mu_i}.
    \label{eq:app_general_concentration}
\end{equation}
The corresponding susceptibility matrix is
\begin{equation}
    X_{ij}^\alpha
    =
    \frac{\partial c_i^\alpha}{\partial \mu_j}
    =
    -\frac{\partial^2\omega^\alpha}
    {\partial\mu_i\partial\mu_j}.
    \label{eq:app_general_susceptibility}
\end{equation}
For a general multicomponent alloy, \(X_{ij}^\alpha\) is not diagonal, and cross thermodynamic couplings may be important. In the dilute independent-solute approximation used here, these couplings are neglected and
\begin{equation}
    X_{ij}^\alpha
    \approx
    X_i^\alpha\delta_{ij}.
    \label{eq:app_diagonal_approximation}
\end{equation}
This diagonalization is the approximation that reduces the general multicomponent thermodynamic description to the dilute model used in this work.

\subsection{Separable dilute phase grand potentials}
\label{app:separable_dilute_gp}

In the diagonal dilute limit, the phase grand-potential density is taken to be separable in the diffusion potentials:
\begin{equation}
    \omega^\alpha(\boldsymbol{\mu},T)
    =
    \omega_0^\alpha(T)
    -
    \sum_i
    \mathcal A_i
    \exp\left[
        \frac{\mu_i-\epsilon_i^\alpha}{\mathcal A_i}
    \right],
    \label{eq:app_phase_gp}
\end{equation}
where \(\omega_0^\alpha(T)\) is the pure-solvent contribution, \(\epsilon_i^\alpha\) is the dilute solute free-energy offset in phase \(\alpha\), and \(\mathcal A_i\) is a component-specific susceptibility parameter.

Using Eq.~\eqref{eq:app_general_concentration}, the phase concentration is
\begin{equation}
    c_i^\alpha
    =
    \exp\left[
        \frac{\mu_i-\epsilon_i^\alpha}{\mathcal A_i}
    \right].
    \label{eq:app_phase_concentration}
\end{equation}
This relation can be inverted as
\begin{equation}
    \mu_i
    =
    \mathcal A_i\ln c_i^\alpha+\epsilon_i^\alpha.
    \label{eq:app_mu_in_terms_c}
\end{equation}
The diagonal susceptibility is
\begin{equation}
    X_i^\alpha
    =
    \frac{\partial c_i^\alpha}{\partial\mu_i}
    =
    \frac{c_i^\alpha}{\mathcal A_i}.
    \label{eq:app_phase_susceptibility}
\end{equation}
Thus \(\mathcal A_i\) is the inverse local concentration susceptibility, scaled by concentration. Unlike the direct ideal dilute extension, \(\mathcal A_i\) is not assumed to be the same for all solutes.

\subsection{Partition coefficients and equilibrium profiles}
\label{app:partition_profiles}

Across the diffuse interface, the solute free-energy offset is interpolated as
\begin{equation}
    \epsilon_i(\phi)
    =
    \bar{\epsilon}_i
    +
    \frac{\Delta\epsilon_i}{2}g(\phi),
    \label{eq:app_epsilon_interp}
\end{equation}
where
\begin{equation}
    \bar{\epsilon}_i
    =
    \frac{\epsilon_i^s+\epsilon_i^l}{2},
    \qquad
    \Delta\epsilon_i
    =
    \epsilon_i^s-\epsilon_i^l.
\end{equation}
The interpolation function satisfies \(g(-1)=-1\) and \(g(+1)=+1\), and \(g'(\pm1)=g''(\pm1)=0\), promoting stable bulk phases at \(\phi=\pm1\).

At bulk equilibrium, equality of diffusion potentials gives
\begin{equation}
    \mathcal A_i\ln c_i^s+\epsilon_i^s
    =
    \mathcal A_i\ln c_i^l+\epsilon_i^l.
\end{equation}
Therefore
\begin{equation}
    k_i
    =
    \frac{c_i^s}{c_i^l}
    =
    \exp\left[
        -\frac{\Delta\epsilon_i}{\mathcal A_i}
    \right],
    \label{eq:app_ki}
\end{equation}
or
\begin{equation}
    \Delta\epsilon_i
    =
    -\mathcal A_i\ln k_i.
    \label{eq:app_delta_epsilon_diag}
\end{equation}
Defining
\begin{equation}
    b_i=\frac{1}{2}\ln k_i,
\end{equation}
the interpolated solute offset can be written as
\begin{equation}
    \epsilon_i(\phi)
    =
    \bar{\epsilon}_i
    -
    \mathcal A_i b_i g(\phi).
    \label{eq:app_epsilon_bi}
\end{equation}

At a stationary equilibrium interface, \(\mu_i\) is spatially uniform. Taking the liquid concentration \(c_i^{l,0}\) as reference gives
\begin{equation}
    c_i^0(\phi)
    =
    c_i^{l,0}
    \exp\left[
        b_i(1+g(\phi))
    \right].
    \label{eq:app_equilibrium_profile}
\end{equation}
This profile has the correct bulk limits:
\begin{equation}
    c_i^0(-1)=c_i^{l,0},
    \qquad
    c_i^0(+1)=k_i c_i^{l,0}.
\end{equation}

\subsection{Liquidus matching and susceptibility parameters}
\label{app:liquidus_matching_susceptibility}

Using Eq.~\eqref{eq:app_phase_gp}, the bulk grand-potential density of phase \(\alpha=s,l\) can be written as
\begin{equation}
    \omega^\alpha
    =
    \omega_0^\alpha(T)
    -
    \sum_i \mathcal A_i c_i^\alpha.
\end{equation}
Bulk coexistence requires
\begin{equation}
    \omega^s=\omega^l.
\end{equation}
Near the pure-solvent melting point,
\begin{equation}
    \omega_0^s(T)-\omega_0^l(T)
    =
    \Delta f_T(T-T_m),
    \label{eq:app_delta_fT_diag}
\end{equation}
where \(\Delta f_T\) is the pure-solvent thermal free-energy coefficient. Substituting this into the coexistence condition and using \(c_i^s=k_i c_i^l\) gives
\begin{equation}
    \Delta f_T(T-T_m)
    +
    \sum_i
    \mathcal A_i(1-k_i)c_i^l
    =
    0.
\end{equation}
Therefore
\begin{equation}
    T-T_m
    =
    -
    \sum_i
    \frac{\mathcal A_i(1-k_i)}{\Delta f_T}
    c_i^l.
    \label{eq:app_liquidus_diag}
\end{equation}
Matching this expression to the prescribed dilute liquidus hypersurface,
\begin{equation}
    T-T_m
    =
    -
    \sum_i m_i c_i^l,
\end{equation}
gives
\begin{equation}
    \mathcal A_i
    =
    \frac{m_i\Delta f_T}{1-k_i}.
    \label{eq:app_Ai_matching}
\end{equation}
At the reference liquid concentration,
\begin{equation}
    X_i^{l,0}
    =
    \frac{c_i^{l,0}}{\mathcal A_i}
    =
    \frac{c_i^{l,0}(1-k_i)}
    {m_i\Delta f_T}.
    \label{eq:app_Xi_matching}
\end{equation}
Thus the independent liquidus slopes are represented by independent diagonal susceptibilities.

\subsection{Thermal interpolation for the reference equilibrium interface}
\label{app:thermal_interpolation}

The local grand-potential density across the diffuse interface is
\begin{equation}
    \omega(\phi,\boldsymbol{\mu},T)
    =
    h f_{\rm dw}(\phi)
    +
    f_T(\phi)(T-T_m)
    -
    \sum_i
    \mathcal A_i
    \exp\left[
        \frac{\mu_i-\epsilon_i(\phi)}{\mathcal A_i}
    \right].
\end{equation}
The variational derivative with respect to \(\phi\), excluding the gradient contribution, contains the thermodynamic driving term
\begin{equation}
    h(\phi^3-\phi)
    +
    f_T'(\phi)(T-T_m)
    +
    \sum_i \epsilon_i'(\phi)c_i.
\end{equation}
The thermal interpolation \(f_T(\phi)\) is chosen so that, at the reference equilibrium interface, the thermal and solutal driving terms cancel pointwise. At the reference composition,
\begin{equation}
    T_m-T=M_0,
    \qquad
    M_0=\sum_i m_i c_i^{l,0},
\end{equation}
and the equilibrium concentration profile is Eq.~\eqref{eq:app_equilibrium_profile}. Thus we require
\begin{equation}
    f_T'(\phi)(T-T_m)
    +
    \sum_i
    \epsilon_i'(\phi)c_i^0(\phi)
    =
    0.
\end{equation}
Using
\begin{equation}
    \epsilon_i'(\phi)
    =
    -\mathcal A_i b_i g'(\phi),
\end{equation}
we obtain
\begin{equation}
    f_T'(\phi)
    =
    -
    \frac{g'(\phi)}{M_0}
    \sum_i
    \mathcal A_i b_i c_i^{l,0}
    \exp\left[
        b_i(1+g(\phi))
    \right].
    \label{eq:app_fT_prime}
\end{equation}
Integrating gives
\begin{equation}
    f_T(\phi)
    =
    -
    \frac{1}{M_0}
    \sum_i
    \mathcal A_i c_i^{l,0}
    \exp\left[
        b_i(1+g(\phi))
    \right].
    \label{eq:app_fT}
\end{equation}
The additive constant is immaterial and set to zero. The jump in \(f_T\) is
\begin{align}
    f_T(+1)-f_T(-1)
    &=
    -\frac{1}{M_0}
    \sum_i
    \mathcal A_i c_i^{l,0}(k_i-1)
    \nonumber\\
    &=
    \frac{1}{M_0}
    \sum_i
    \mathcal A_i(1-k_i)c_i^{l,0}.
\end{align}
Using Eq.~\eqref{eq:app_Ai_matching} and \(M_0=\sum_i m_i c_i^{l,0}\), this becomes
\begin{equation}
    f_T(+1)-f_T(-1)
    =
    \Delta f_T.
\end{equation}
Thus the thermal interpolation used in the diffuse-interface model is consistent with the liquidus matching.

\section{Sharp-interface matching of the phase field equations}
\label{app:sharp_interface_matching}

This section derives the parameter relations used to match the phase field equations to the sharp-interface Gibbs-Thomson condition. The calculation is performed for the isotropic one-dimensional interface. The anisotropic multidimensional form used in simulations is obtained by replacing \(W\) and \(\tau_0\) by their orientation-dependent counterparts as described in the main text.

\subsection{Order-parameter and concentration equations}
\label{app:evolution_equations_derivation}

The grand potential functional is
\begin{equation}
    \Omega[\phi,\boldsymbol{\mu},T]
    =
    \int
    \left[
        \frac{\sigma}{2}|\nabla\phi|^2
        +
        \omega(\phi,\boldsymbol{\mu},T)
    \right]dV.
\end{equation}
The non-conserved dynamics of the order parameter are obtained from the variational derivative
\begin{equation}
    \frac{\partial\phi}{\partial t}
    =
    -K_\phi\frac{\delta\Omega}{\delta\phi}.
\end{equation}
With
\begin{equation}
    \tau_0=\frac{1}{K_\phi h}, \qquad \sigma=hW^2
\end{equation}
and using Eq.~\eqref{eq:app_fT_prime}, the concentration-variable form of the isotropic phase field equation follows as
\begin{equation}
    \tau_0\frac{\partial\phi}{\partial t}
    =
    W^2\nabla^2\phi+\phi-\phi^3
    -
    g'(\phi)
    \sum_i
    \lambda_i
    \left[
        c_i
        +
        \frac{T-T_m}{M_0}
        c_i^{l,0}
        e^{b_i(1+g(\phi))}
    \right],
    \label{eq:app_phi_eq}
\end{equation}
where
\begin{equation}
    \lambda_i
    =
    -\frac{\mathcal A_i b_i}{h}.
    \label{eq:app_lambda_Ai}
\end{equation}

The concentration dynamics follow conserved dynamics, written as 
\begin{equation}
    \frac{\partial c_i}{\partial t}
    =
    \nabla\cdot
    \left[
        K_i\nabla\mu_i
    \right].
\end{equation}
Choosing
\begin{equation}
    K_i
    =
    \frac{D_l^i q_i(\phi)c_i}{\mathcal A_i}
\end{equation}
and using
\begin{equation}
    \mu_i
    =
    \mathcal A_i\ln c_i+\epsilon_i(\phi)
    =
    \mathcal A_i
    \left[
        \ln c_i-b_i g(\phi)
    \right]
\end{equation}
gives
\begin{equation}
    \frac{\partial c_i}{\partial t}
    =
    \nabla\cdot
    \left\{
        D_l^i q_i(\phi)c_i
        \nabla
        \left[
            \ln c_i-b_i g(\phi)
        \right]
    \right\},
    \label{eq:app_concentration_eq}
\end{equation}
ensuring Fickian diffusion in the bulk liquid, and no diffusion in the bulk solid when \(q(-1)=+1\), and \(q(+1)=0\), respectively.

\subsection{Integrated order-parameter solvability condition}
\label{app:integrated_solvability}

Consider a steady interface moving with velocity \(V\) in the normal direction. In the moving frame,
\begin{equation}
    \partial_t\phi
    =
    -V\frac{d\phi}{dx}.
\end{equation}
Including curvature through
\begin{equation}
    \nabla^2\phi
    =
    \frac{d^2\phi}{dx^2}
    +
    \kappa\frac{d\phi}{dx},
\end{equation}
Eq.~\eqref{eq:app_phi_eq} becomes
\begin{align}
    -\tau_0 V\frac{d\phi}{dx}
    &=
    W^2
    \left(
        \frac{d^2\phi}{dx^2}
        +
        \kappa\frac{d\phi}{dx}
    \right)
    +
    \phi-\phi^3
    \nonumber\\
    &\quad
    -
    g'(\phi)
    \sum_i
    \lambda_i
    \left[
        c_i
        +
        \frac{T-T_m}{M_0}
        c_i^{l,0}
        e^{b_i(1+g(\phi))}
    \right].
    \label{eq:app_moving_phi}
\end{align}
Multiplying by \(d\phi/dx\) and integrating through the interface gives the solvability condition. The terms involving \(d^2\phi/dx^2\) and \(\phi-\phi^3\) vanish.

We define
\begin{equation}
    a_1
    =
    W
    \int_{-\infty}^{+\infty}
    \left(
        \frac{d\phi}{dx}
    \right)^2 dx,
    \label{eq:app_a1_def}
\end{equation}
and \(\phi_0(x)=-\tanh[x/(\sqrt{2}W)]\) as the equilibrium planar profile. For this equilibrium profile Eq.~\eqref{eq:app_a1_def} gives,
\begin{equation}
    a_1^0=\frac{2\sqrt{2}}{3}.
\end{equation}
The velocity and curvature terms give
\begin{equation}
    \int
    -\tau_0 V
    \left(
        \frac{d\phi_0}{dx}
    \right)^2 dx
    =
    -\tau_0 V\frac{a_1^0}{W},
\end{equation}
and
\begin{equation}
    \int
    W^2\kappa
    \left(
        \frac{d\phi_0}{dx}
    \right)^2 dx
    =
    a_1^0 W\kappa,
\end{equation}
respectively. We define the nonequilibrium solute integral
\begin{equation}
    I_i(V)
    =
    \int_{-\infty}^{+\infty}
    g'(\phi_0)c_i
    \frac{d\phi_0}{dx}\,dx,
    \label{eq:app_Ii}
\end{equation}
and its equilibrium counterpart
\begin{equation}
    I_i^0
    =
    \int_{-\infty}^{+\infty}
    g'(\phi_0)c_i^{l,0}
    e^{b_i(1+g(\phi_0))}
    \frac{d\phi_0}{dx}\,dx.
\end{equation}
Changing variables from \(x\) to \(\phi_0\), and using \(\phi_0(-\infty)=+1\), \(\phi_0(+\infty)=-1\), gives
\begin{align}
    I_i^0
    &=
    c_i^{l,0}
    \int_{+1}^{-1}
    g'(\phi_0)
    e^{b_i(1+g(\phi_0))}\,d\phi
    \nonumber\\
    &=
    \frac{c_i^{l,0}(1-k_i)}{b_i}.
    \label{eq:app_Ii0}
\end{align}

Collecting the integrated terms gives
\begin{equation}
    -\tau_0 V\frac{a_1^0}{W}
    =
    a_1^0W\kappa
    -
    \sum_i\lambda_i I_i(V)
    -
    \frac{T-T_m}{M_0}
    \sum_i\lambda_i I_i^0 .
\end{equation}
Equivalently,
\begin{equation}
    -\tau_0 V\frac{a_1^0}{W}
    =
    a_1^0W\kappa
    -
    \sum_i\lambda_i I_i(V)
    +
    (T_m-T)\Lambda,
    \label{eq:app_integrated_raw}
\end{equation}
where
\begin{equation}
    \Lambda
    =
    \frac{1}{M_0}
    \sum_i\lambda_i I_i^0 .
    \label{eq:app_Lambda_def}
\end{equation}
Solving Eq.~\eqref{eq:app_integrated_raw} for \(T\) gives
\begin{equation}
    T
    =
    T_m
    -
    \sum_i
    \frac{\lambda_i}{\Lambda}
    I_i(V)
    +
    \frac{a_1^0W}{\Lambda}\kappa
    +
    \frac{\tau_0a_1^0}{\Lambda W}V.
    \label{eq:app_temperature_response}
\end{equation}

Using Eqs.~\eqref{eq:app_lambda_Ai} and \eqref{eq:app_Ii0},
\begin{align}
    \lambda_i I_i^0
    &=
    \left(
        -\frac{\mathcal A_i b_i}{h}
    \right)
    \frac{c_i^{l,0}(1-k_i)}{b_i}
    \nonumber\\
    &=
    -\frac{\mathcal A_i(1-k_i)c_i^{l,0}}{h}.
\end{align}
Using \(\mathcal A_i(1-k_i)=m_i\Delta f_T\), one obtains
\begin{equation}
    \Lambda
    =
    -\frac{\Delta f_T}{h}.
    \label{eq:app_Lambda_delta_fT}
\end{equation}

\subsection{Capillary and kinetic matching}
\label{app:capillary_kinetic_matching}

The desired sharp-interface Gibbs-Thomson relation is
\begin{equation}
    T
    =
    T_m
    -
    \sum_i m_i(V)c_i^l
    -
    \Gamma\kappa
    -
    \frac{V}{\mu_0},
    \label{eq:app_GT_target}
\end{equation}
where \(\Gamma\) is the Gibbs-Thomson coefficient and \(\mu_0\) is the reference kinetic coefficient.

Comparing the curvature term in Eq.~\eqref{eq:app_temperature_response} with Eq.~\eqref{eq:app_GT_target} gives
\begin{equation}
    \frac{a_1^0W}{\Lambda}
    =
    -\Gamma.
\end{equation}
Using Eq.~\eqref{eq:app_Lambda_delta_fT}, this yields
\begin{equation}
    \Delta f_T
    =
    \frac{h a_1^0 W}{\Gamma}.
    \label{eq:app_delta_fT_matching}
\end{equation}
Matching the kinetic term gives
\begin{equation}
    \frac{\tau_0a_1^0}{\Lambda W}
    =
    -\frac{1}{\mu_0}.
\end{equation}
Using \(\Lambda=-a_1^0W/\Gamma\), this gives
\begin{equation}
    \tau_0
    =
    \frac{W^2}{\Gamma\mu_0}.
    \label{eq:app_tau_matching}
\end{equation}
Finally, substituting Eq.~\eqref{eq:app_delta_fT_matching} into
\begin{equation}
    \lambda_i
    =
    -\frac{\mathcal A_i b_i}{h}
    =
    -\frac{m_i\Delta f_T}{1-k_i}
    \frac{b_i}{h}
\end{equation}
gives
\begin{equation}
    \lambda_i
    =
    -\frac{a_1^0W}{\Gamma}
    \frac{m_i b_i}{1-k_i}.
    \label{eq:app_lambda_final}
\end{equation}

\subsection{Velocity-dependent liquidus response}
\label{app:velocity_dependent_liquidus_response}

The compositional term in Eq.~\eqref{eq:app_temperature_response} must match the corresponding term in Eq.~\eqref{eq:app_GT_target}. Therefore
\begin{equation}
    m_i(V)c_i^l
    =
    \frac{\lambda_i}{\Lambda}I_i(V).
\end{equation}
Using
\begin{equation}
    \frac{\lambda_i}{\Lambda}
    =
    \frac{m_i b_i}{1-k_i},
\end{equation}
we obtain
\begin{equation}
    \frac{m_i^{\rm PF}(V)}{m_i}
    =
    \frac{b_i}{(1-k_i)c_i^l}
    \int_{-\infty}^{+\infty}
    g'(\phi)c_i
    \frac{d\phi}{dx}\,dx.
    \label{eq:app_miV_final}
\end{equation}
In equilibrium, \(c_i\rightarrow c_i^0(\phi)\), \(c_i^l\rightarrow c_i^{l,0}\), and Eq.~\eqref{eq:app_miV_final} gives \(m_i^{\rm PF}(0)=m_i\).

\section{Chebyshev polynomial representation}
\label{sec:supp_chebyshev}

\begin{figure*}[ht!]
    \centering
    \includegraphics[]{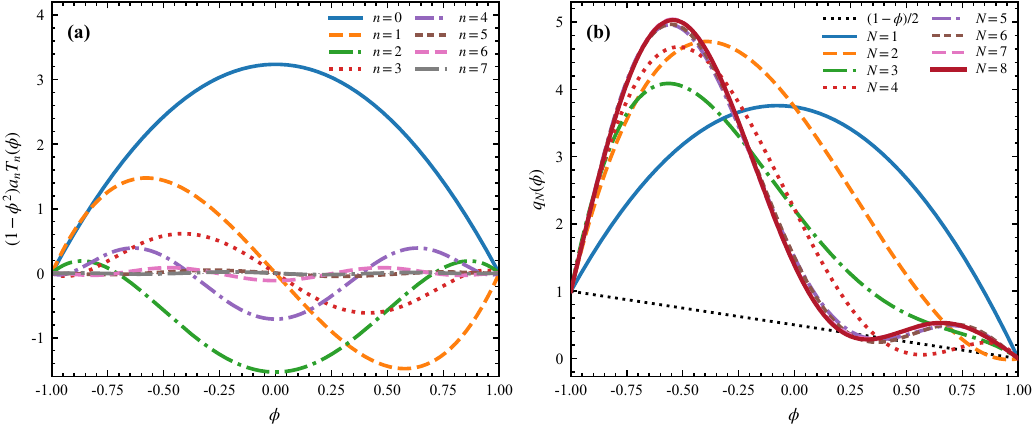}
    \caption{
    Construction of the optimized Chebyshev representation of the
    interfacial diffusivity for the binary case with \(\alpha=0.3\).
    (a) Individual weighted contributions
    \((1-\phi^2)a_nT_n(\phi)\) of the eight optimized basis functions.
    (b) Cumulative representations obtained by successively including
    the optimized coefficients in Eq.~\eqref{eq:supp_chebyshev_representation}.
    The higher-order terms produce progressively smaller changes in the
    resulting \(q(\phi)\), supporting the use of \(N=8\) in the
    optimization.
    }
    \label{fig:supp_chebyshev_convergence}
\end{figure*}

The interfacial diffusivity is represented as \begin{equation} 
    q_N(\phi) = \frac{1-\phi}{2} + (1-\phi^2) \sum_{n=0}^{N-1} a_n T_n(\phi), 
\label{eq:supp_chebyshev_representation} 
\end{equation} 
where \(T_n(\phi)\) denotes the Chebyshev polynomial of degree \(n\). The prefactor \(1-\phi^2\) ensures that the correction vanishes in the bulk phases, independently of the expansion order. Figure~\ref{fig:supp_chebyshev_convergence} illustrates the construction of the optimized \(N=8\) representation for the binary case with \(\alpha=0.3\). The individual weighted terms \((1-\phi^2)a_nT_n(\phi)\) are shown in Fig.~\ref{fig:supp_chebyshev_convergence}(a), while Fig.~\ref{fig:supp_chebyshev_convergence}(b) shows the cumulative representation obtained by successively including the optimized coefficients. The first few terms determine the principal shape of \(q(\phi)\), whereas the higher-order terms introduce only progressively smaller refinements. The cumulative representation is therefore effectively established by \(N=8\) which was consequently used throughout the calculations presented in the main article.

\section{Grid-resolution check for the two-dimensional calculations}

A grid-resolution check was performed using a moving-frame directional-solidification calculation initialized from a circular solid seed. The pulling direction was along \(x\), with the dendrite tip maintained near the center of the \(x\)-dimension by repositioning the moving frame. The moving-frame domain size was \(L_x=5~\mu\mathrm{m}\) and \(L_y=0.72~\mu\mathrm{m}\). The same Al--Cu material parameters were used as in the main text. The imposed pulling velocity and thermal gradient were \(V_{\rm pull}=0.1~\mathrm{m\,s^{-1}}\) and \(G=5\times 10^6~\mathrm{K\,m^{-1}}\), respectively, and the simulation was run for \(100~\mu\mathrm{s}\). The transverse length \(L_y=0.72~\mu\mathrm{m}\) was chosen so that the tested grid spacings are commensurate with the domain, where \(W=SW_0=5~\mathrm{nm}\) and the tested grid spacings are \(\Delta x/W=0.4\), \(0.6\), \(0.8\), and \(1.0\).

\begin{figure}[ht!]
    \centering
    \includegraphics[]{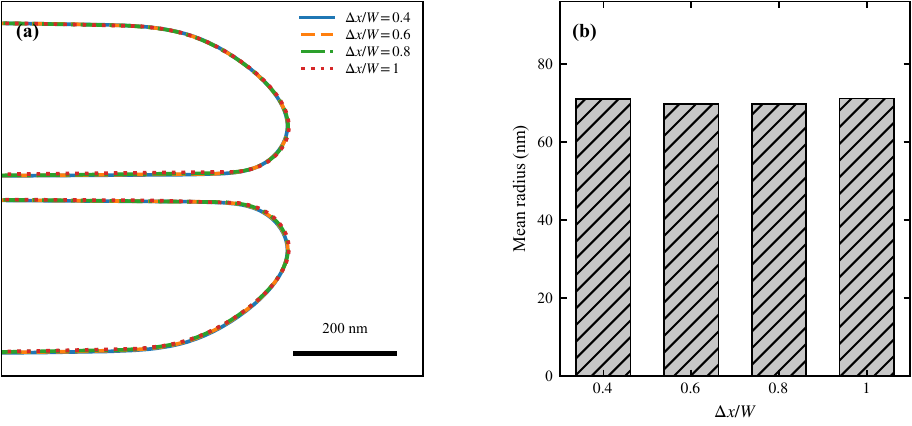}
    \caption{
    Grid-resolution check for the two-dimensional dendrite calculation. 
    (a) Extracted \(\phi=0\) fronts after alignment for different values of \(\Delta x/W\).
    (b) Mean fitted tip radius as a function of grid spacing.
    }
    \label{fig:supp_split_tip_convergence}
\end{figure}

During the benchmark, the initially circular seed developed a split tip morphology illustrated in Figure~\ref{fig:supp_split_tip_convergence}. For each resolution, the front position \(x_{\mathrm f}(y)\) was obtained by one-dimensional interpolation of the \(\phi=0\) crossing along the growth direction. The fronts were then aligned from the separate simulations and compared in physical coordinates. The close overlap of the fronts in Fig.~\ref{fig:supp_split_tip_convergence}(a) indicates that the resolved morphology is only weakly affected by the grid spacing over the range tested.

To obtain a compact curvature metric, the two tip regions on either side of the central liquid channel were fitted separately. In each case, local points on the extracted \(\phi=0\) front were fitted to a parabola,
\begin{equation}
    x(y)=a(y-y_t)^2+b(y-y_t)+c,
\end{equation}
over a fixed physical depth behind the tip. The local radius was then computed from the fitted curvature as
\begin{equation}
    R=\frac{(1+b^2)^{3/2}}{2|a|}.
\end{equation}
The reported radius for each grid spacing is the average of the two fitted tip radii at a selected fitting depth of \(5~\mathrm{nm}\) corresponding to 8--24 grid points used in each fit. This mean radius varies only modestly across the tested resolutions as shown in Figure~\ref{fig:supp_split_tip_convergence}(b). Based on the agreement of the front shapes and fitted radii, \(\Delta x/W=0.8\) was selected for the larger production simulations as a compromise between numerical accuracy and computational cost.

\end{document}